\documentclass[fleqn,usenatbib]{mnras}

\usepackage{newtxtext,newtxmath}
\usepackage[T1]{fontenc}

\DeclareRobustCommand{\VAN}[3]{#2}
\let\VANthebibliography\thebibliography
\def\thebibliography{\DeclareRobustCommand{\VAN}[3]{##3}\VANthebibliography}

\usepackage{graphicx}
\usepackage{amsmath}	%

\title[]{Much Ado About Nothing: Galaxy Formation and Galactic Outflows in Cosmic Voids}

\author[Pan et al.]{
Yue Pan,$^{1}$\thanks{E-mail: yue.pan@princeton.edu}
Romain Teyssier,$^{1}$
Ulrich P. Steinwandel, $^{2,3}$
Alice Pisani$^{1,4}$
\\
$^{1}$Department of Astrophysical Sciences, Princeton University, Peyton Hall, Princeton, NJ 08544, USA\\
$^{2}$Center for Computational Astrophysics, Flatiron Institute, 162 5th Ave., New York, NY 10010, USA\\
$^{3}$Max-Planck-Institut für Astrophysik, Karl-Schwarzschild-Straße 1, 85741 Garching, Germany\\
$^{4}$Aix-Marseille Université, CNRS/IN2P3, CPPM, Marseille, France
}

\date{Accepted XXX. Received YYY; in original form ZZZ}

\pubyear{2024}

\begin{document}
\label{firstpage}
\pagerange{\pageref{firstpage}--\pageref{lastpage}}
\maketitle

\begin{abstract}
We present a theoretical framework for calculating the volume filling fraction of galactic outflows in cosmic voids by integrating analytical models for the halo mass function (HMF), the halo occupation fraction, the stellar mass–halo mass relation, and outflow sizes. Using \texttt{RAMSES}, we perform a hydrodynamical zoom-in simulation of the central 25 cMpc/h region of a spherical void, identified as the lowest-density region among 1,000 random spheres in a parent 1 Gpc box simulation. This void has a diameter of 120 cMpc/h and a density contrast of $\delta \simeq -0.8$. We find that the properties of void galaxies remain stable when expanding the zoom-in region to 50 cMpc/h, though our relatively low mass resolution impacts the results. Our higher-resolution simulation aligns with the analytical HMF that accounts for the void’s underdensity and size. While higher resolution improves stellar mass estimates for low-mass halos, computational constraints necessitate a theoretical framework that enables extrapolation to infinite resolution. Our analytical model, calibrated to our simulations, enables extrapolation down to the filtering mass of star-forming halos.
To compare galaxy properties in this void with those in the field, we conduct a companion field simulation of the same box size. At infinite resolution, we predict wind volume filling fractions of $18.6\%$ in the field and $3.1\%$ in our void, with values dependent on cosmic variance, void size, and underdensity. Dwarf galaxies contribute minimally, and resolving halos to $M_{\rm h}=10^{10} M_\odot$ suffices for robust estimates. Applying our framework to the Local Group void ($\delta \simeq -0.5$, $R=20\ \mathrm{cMpc}$), we predict a wind volume filling fraction of $9.6\%\pm3.3\%$.

\end{abstract}



\section{Introduction}
Much of the mass in the Universe is confined within virialized structures such as galaxies, clusters, and filaments. However, the vast majority of the volume is occupied by large underdense voids, which form, in that sense, the dominant component of the large-scale galaxy and matter distributions. These cosmic voids, spanning tens to hundreds of megaparsecs in diameter, are regions with significantly lower densities than the cosmic mean and play a critical role in shaping the topology of the Universe. In a void-centric description of structure formation, matter is redistributed as voids expand, squeezing into sheets and filaments that form at the intersections of void walls \citep{Icke.1984, van.de.Weygaert.1993}. This void-driven redistribution process shapes the cosmic web, making voids essential to understanding the evolution of large-scale structure.

Voids arise in regions with underdense initial conditions, where peculiar velocities create an effective repulsion, driving the expansion of the void. This leads to shell-crossing events, where the inner shells of matter overtake the outer shells, resulting in self-similar density profiles \citep{Fillmore.Goldreich.1984, Sheth.van.de.weygaert.2004, Hamaus.etal.2014, Nadathur.etal.2015}. The simple dynamics of void evolution make them particularly appealing for theoretical and observational studies, as their profiles often exhibit universal behavior that does not depend on their size \citep{Shim.etal.2021}. Unlike overdense regions, which collapse into complex structures, voids evolve toward greater symmetry, making them more predictable and accessible to analysis \citep{Icke.1984, Bertschinger.1985, van.de.Weygaert.1993, Hamaus.etal.2014}.

In recent years, voids have emerged as powerful cosmological probes. When combined with traditional galaxy-based methods, void statistics provide tighter constraints on key cosmological parameters. For instance, combining the halo mass function (HMF), the non-linear matter power spectrum, and the void size function (VSF) can break degeneracies in cosmological models \citep{Bayer.etal.2021, Kreisch.etal.2022, Pelliciari.etal.2023, Contarini.etal.2023}. Similarly, void clustering offers competitive constraints on parameters such as the Hubble constant ($\rm h_{0}$) and baryon density ($\Omega_{\rm b}$). Including void-lensing cross-correlation signals improves constraints on the spectral index of primordial fluctuations ($n_{\rm s}$) by 10–15\% \citep{Bonici.etal.2023}. The sensitivity of voids to the geometry of the Universe and the properties of dark energy is particularly strong. For example, the void size function and shapes have been used, to constrain the equation of state of dark energy, significantly tightening the $w_0-w_{\rm a}$ parameter space when combined with other probes \citep{Pisani.etal.2015, Verza.etal.2019}.

Void-based studies in upcoming surveys such as \emph{Euclid} promise to revolutionize cosmology by exploiting voids as standalone probes. For instance, the combination of void size function and void-galaxy cross-correlation signals in BOSS and \emph{Euclid} is expected to constrain the dark energy equation of state with a precision of 10\% \citep{Hamaus.etal.2021, Contarini.etal.2022}. Constraints from voids measured in denser surveys will provide further complementary information on these cosmological parameters \citep[e.g., Nancy Grace Roman Space Telescope,][]{Verza.etal.2024b}. Additionally, joint analyses of geometric and dynamical distortions in void shapes are predicted to provide competitive constraints on $\Omega_m$ and cosmological distance ratios, with errors on growth rate parameters reduced by 5–8\% \citep{Hamaus.etal.2022}. Voids are also sensitive to neutrino masses, with studies forecasting a gain of up to 60\% in precision when voids are combined with halos \citep{Kreisch.etal.2019, Kreisch.etal.2022}.

Beyond their cosmological significance, voids are also unique laboratories for studying galaxy formation and evolution. Galaxies in voids tend to exhibit distinct properties compared to those in denser environments, including later morphological types, lower masses, and slower evolutionary rates \citep{Conrado.etal.2024, Elyiv.etal.2013}. Star formation in void galaxies is often more efficient relative to their stellar mass, particularly closer to void centers, because they experience delayed growth in underdense environments, retain their gas more effectively due to weaker external pressures, and avoid strong quenching mechanisms such as ram pressure stripping and frequent mergers, which are prevalent in denser cosmic regions \citep{Habouzit.etal.2020}. These findings suggest that the low-density environments of voids fundamentally alter the processes of galaxy evolution, providing insights that cannot be gleaned from studies of overdense regions alone.

Another fascinating aspect of voids is their density profiles, which exhibit remarkable self-similarity across scales. This universality enables the use of void density profiles as robust tools for testing theories of structure formation. Studies have shown that void profiles are well described by simple, parametric models that capture the effects of expansion and shell-crossing \citep{Fillmore.Goldreich.1984, Shim.etal.2021}. Moreover, advancements in void identification techniques, such as the topological ZOBOV void finder \citep{Platen.etal.2007, Neyrinck.2008}, have enabled the creation of detailed void catalogs that span diverse environments and redshifts.

One of the key distinctions between voids and other cosmic environments lies in the efficiency of galaxy formation. The low-density conditions of voids lead to delayed structure growth, resulting in galaxies that tend to be lower in mass, bluer, and more actively star-forming compared to their counterparts in denser regions \citep{Conrado.etal.2024, Elyiv.etal.2013}. However, the efficiency of star formation is not solely dictated by gas availability – it is also influenced by feedback processes, particularly galactic winds, which regulate star formation by redistributing gas within and beyond galaxies. These winds, driven by stellar feedback and active galactic nuclei, can alter the local gas density and influence the extent to which galaxies pollute their surroundings. Voids, with their minimal contamination by galactic outflows, serve as unique environments to explore how these feedback-driven outflows operate compared to denser regions. 

Additionally, voids act as reservoirs of pristine primordial gas, providing a natural setting to investigate the imprint of primordial magnetic fields \citep{Beck.etal.2013, Rodriguez-Medrano.etal.2023}. The unique low-density and low-metallicity environments of voids make them compelling regions for studying early star formation processes. These conditions are thought to be conducive to the formation of massive, metal-free Population III stars. Recent studies \citep[e.g.][]{Ricciardelli.etal.2014} have suggested that void galaxies exhibit distinct star formation activities compared to those in denser regions, potentially retaining more pristine conditions that could be analogous to the environments where Population III stars formed.

In this paper, we investigate the extent to which voids are polluted by metals from galaxy formation and galactic winds. Understanding this process is essential for assessing how pristine voids truly are, as their composition depends critically on the physics of galaxy formation and the specific characteristics of galaxies residing within them. To quantify this effect, we introduce the \textit{galactic wind volume filling fraction}, defined as the ratio of the total volume occupied by wind-blown material from all galaxies in the void to the total simulation volume. This metric serves as a key tool for probing the impact of galactic winds on void environments. By analyzing the volume filling fraction, we can assess how the redistribution of matter through galactic winds influences both the internal structure of voids and the surrounding cosmic web. Leveraging the simplicity of void dynamics and their unique characteristics, we aim to deepen our understanding of how voids contribute to shaping the large-scale structure of the Universe. Through this study, we explore their potential as valuable cosmological and astrophysical laboratories.

The structure of this paper is organized to explore various facets of galaxy formation and matter distribution in cosmic voids. In Sec.~\ref{sec:sim}, we describe the simulation setup, including the identification and characterization of void regions. In Sec.~\ref{sec:halo_stats} we delve into halo statistics, presenting a theoretical framework for the halo mass function, halo occupation fraction, and stellar mass--halo mass relation, and their dependence on environmental properties. The dynamics of galaxy formation and stellar feedback within voids are detailed in Sec.~\ref{sec:wind}, where we also compare these processes to field environments. In Sec.~\ref{sec:discussions}, we discuss the volume filling fraction of galaxies in different environments, connecting theoretical predictions with simulation results. Finally, Sec.~\ref{sec:summary} highlights the main results and outlines future research directions based on the insights gained from this work.

\section{Simulation setup}\label{sec:sim}

\subsection{RAMSES overview} \label{sec:ramses}
We run a \texttt{RAMSES}\footnote{\url{https://bitbucket.org/rteyssie/ramses/src/master/}} \citep{Teyssier.2002} simulation in a $L_{\rm box} = 1000$ cMpc/$h_0$ cube, employing a $\Lambda$CDM cosmology characterized by a total matter density $\Omega_{\rm m,0} = 0.308$, dark energy density $\Omega_{\Lambda,0} = 0.692$, baryon density $\Omega_{\rm b} = 0.0484$, amplitude of the matter power spectrum $\sigma_8 = 0.8149$, Hubble constant $H_0 = 67.81 \, \rm km \, s^{-1} \, Mpc^{-1}$, and spectral index $n_s = 0.9677$, consistent with the Planck 2018 results \citep{Planck.2018}.

In our simulation setup, gas heating from a uniform UV background is applied following reionization at $z_{\mathrm{reion}} = 10$, as outlined by \citet{Haardt.Madau.1996}. Gas cooling processes down to $10^4 \, \mathrm{K}$ occur via Hydrogen (H) and Helium (He) collisions, with additional contributions from metals using rates provided by \citet{Dopita.Sutherland.1996}. Cooling at lower temperature is also included using a simple model calibrated on Cloudy \citep{Ferland.etal.1998}. Star formation is triggered in regions where the gas number density exceeds $n_0 = 0.1 \, \mathrm{H \, cm^{-3}}$, following a Schmidt relation: $\dot{\rho}_{\ast} = \epsilon_{\ast} \rho_g / t_{\mathrm{ff}}$, where $\dot{\rho}_{\ast}$ represents the star formation rate density, $\rho_g$ is the gas mass density, $\epsilon_{\ast} = 0.02$ is the star formation efficiency per free-fall time, and $t_{\mathrm{ff}}$ denotes the gas free-fall time. To accurately model stellar feedback, the simulation incorporates mass, energy, and metal releases from stellar winds and Type Ia and core collapse SNe.

Furthermore, the simulation includes the formation and growth of black holes (BHs), which can accrete gas at a Bondi rate capped at the Eddington limit and merge if they form a sufficiently close binary. Energy is released by BHs in two modes: a heating (or "quasar") mode when the accretion rate exceeds one percent of the Eddington limit, and a jet (or "radio") mode when it falls below this threshold. Each mode's efficiency is calibrated to match observed BH--galaxy scaling relations at $z = 0$ \citep{Dubois.etal.2012}.

The grid resolution is defined in terms of a refinement level $\ell$, with corresponding cell size $\Delta x = L/2^\ell$, where $L$ is the box length. The mass resolution for dark matter particles is given by $m_{\rm dm, min} = \Omega_{\rm m, 0} \, \rho_{\rm crit, 0} \, (L/2^\ell)^3$, where $\rho_{\rm crit, 0} = 3H_0^2 / (8 \pi G)$ is the critical density of the universe at $z = 0$. The total volume contains $1024^3$ dark matter particles. To increase resolution where needed, the initially coarse $1024^3$ grid is adaptively refined down to $\Delta x$ using a quasi-Lagrangian refinement criterion: if the number of dark matter particles within a cell exceeds 8, or if the total baryonic mass reaches eight times the initial dark matter mass resolution, further refinement is triggered.

Initial conditions for a low-resolution $\ell = 9$ dark matter-only simulation were generated in a comoving volume of $L^3 = (1000 \, \rm cMpc/h_0)^3$ at $z = 124$ using \texttt{MUSIC}\footnote{\url{https://bitbucket.org/ohahn/music/src/master/}} \citep{Hahn.Abel.2011}. This simulation was evolved until $z = 0$, and the final snapshot was used as a parent simulation to identify voids within the volume.

To explore different environments, we also ran a set of dark matter-only and hydrodynamical simulations using the same initial conditions generated from \texttt{MUSIC}, but within a smaller comoving volume of $L^3 = (25 \, \rm cMpc/h_0)^3$ at $z = 124$. Referred to as the "field simulation" in subsequent analysis, this set of simulations was conditioned to have an average density equal to $\rho_{\rm crit, 0}$ (i.e., an overdensity $\delta = 0$). This reference simulation allows us to assess the impact of environment on galaxy properties and galactic winds.

Finally, our simulations were conducted on the Stellar cluster at Princeton University. Each simulation was executed on 4 nodes, with 96 cores per node, totaling 384 cores. The cores on Stellar are 2.9 GHz Intel Cascade Lake processors. The high-resolution void hydrodynamical simulation required approximately 3 weeks to complete, while all other simulations finished in under a week.

\subsection{Identify a void with random spheres} \label{sec:find_void}
The first step, involves identifying a void region within the dark matter only parent simulation box of size 1000$\, \mathrm{cMpc}/\rm{h_0}$. We initially define the void size to have a diameter of 25$\, \mathrm{cMpc}/\rm{h_0}$ to compare with the 25$\, \mathrm{cMpc}/\rm{h_0}$ field simulation. Our particular choice for a box size of 25$\, \mathrm{cMpc}/\rm{h_0}$ is a compromise between the size of the simulated volume and the grid resolution. The CAMELS simulations\footnote{\url{https://www.camel-simulations.org/}} \citep{CAMELS_presentation, CAMELS_DR1, CAMELS_DR2} are based on a similar compromise and have many similar field simulations with box size 25$\, \mathrm{cMpc}/\rm{h_0}$, so our results of the void and the field could be easily compared to these.

To identify a void in the final snapshot of the parent simulation, we randomly select 1000 spheres with a diameter of 25$\, \mathrm{cMpc}/\rm{h_0}$ and calculate their average densities. For spheres that extend beyond the simulation box boundary, we compute the density using the mirrored positions of particles, as periodic boundary conditions were applied in the simulation setup. A spherical region is preferred over a cubic one because it is more isotropic and symmetric.

We select the sphere with the lowest density as our void. It is important to emphasize that this is ``a'' void, not necessarily ``the'' void. The aim of this project is to compare galaxy and galactic wind properties in one cosmically underdense region with those in a more typical environment (like the field simulation), rather than selecting the absolute most underdense 25$\, \mathrm{cMpc}/\rm{h_0}$ sphere in the 1000$\, \mathrm{cMpc}/\rm{h_0}$ dark matter only parent simulation. At this resolution ($\ell = 9$), the lowest-density region primarily reflects the initial filamentary structure generated by \texttt{MUSIC}, shaped by subsequent cosmic expansion and nonlinear structure formation. In contrast, the densest region of the simulation contains two large galaxy clusters.

\subsection{Void properties} \label{sec:void_prop}

\begin{figure}
    \centering
    \includegraphics[width=\columnwidth]{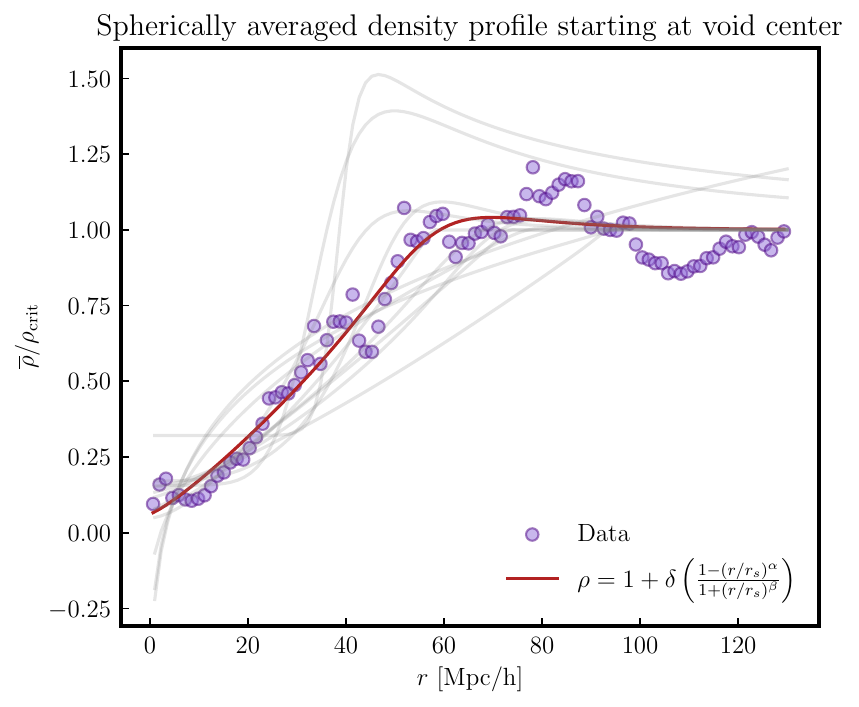}
    \caption{Spherically averaged density profile of our fiducial void. The general trend – extremely low at the center, rising steadily, overshooting above the critical density, and finally plateauing to the critical density – agrees with previous studies, as there is a pileup of structures at the void edge. We fit the model in Equation~\ref{eqn:void_profile} \citep{Hamaus.etal.2014, Nadathur.etal.2015} to our density profile (maroon line) and found good agreement. We also overplot the fitted density profiles for the 12 solid angle bins (Section~\ref{sec:void_prop}) as light grey lines in the background to show the full range of void radii at different directions. }
    \label{fig:void_density}
\end{figure}

Voids are, by definition, underdense regions enclosed by an overdense boundary. They exhibit "universality" \citep{Hamaus.etal.2014}, meaning their density profile is characterized by a low value within the void, an overshoot at the boundary (as matter and structures accumulate at the edges), and a plateau at the mean density of the universe beyond the void boundary \citep{Fillmore.Goldreich.1984, Nadathur.etal.2015, Hamaus.etal.2014}. Figure~\ref{fig:void_density} presents the spherically averaged density profile of the void selected using the random sphere method described in Sec.~~\ref{sec:find_void}.

Since voids are highly aspherical \citep[see e.g.][]{Sutter.etal.2014, Sutter.etal.2015}, determining their precise shape, size, and center is challenging. Our simplified method to estimate the void center involves drawing 1000 spheres from the last snapshot with a fixed diameter of 25$\, \mathrm{cMpc}/\rm{h_0}$ and random centers restricted to the spherical region 2.5$\, \mathrm{cMpc}/\rm{h_0}$ away from the center identified in Sec.~\ref{sec:find_void}. The center of the sphere with the lowest mean density is then taken as our estimate of the new void center, which is close to the center computed in Sec.~\ref{sec:find_void}. We subsequently grow concentric shells around this new void center and compute the spherically averaged density profile of the void.

The spherically averaged density profile in Fig.~\ref{fig:void_density} indicates that the actual size of the void has a radius of approximately 60$\, \mathrm{cMpc}/\rm{h_0}$ (or a diameter of 120$\, \mathrm{cMpc}/\rm{h_0}$), if we define the void's edge as the point where the average density overshoots the mean density of the Universe due to the accumulation of structures at the boundary. We fit the following equation from \citet{Nadathur.etal.2015, Hamaus.etal.2014}:
\begin{equation}\label{eqn:void_profile}
    \frac{\rho(r)}{\bar{\rho}} = 1 + b \left(\frac{1 - (r/r_s)^{\alpha}}{1 + (r/r_s)^{\beta}}\right)
\end{equation}
to our data with parameters $b = -0.94, \alpha = 1.20, \beta =8.35, r_s = 0.06$ and found good agreement. According to Fig.~\ref{fig:void_density}, the mean density $\bar{\rho}/\rho_{\rm crit}$ within the void at a radius of approximately 12.5$\, \mathrm{cMpc}/\rm{h_0}$ is 0.2, which implies that the density contrast of our void is $\delta = \bar{\rho}/\rho_{\rm crit} - 1 = -0.8$. We choose to evaluate the density at $r = 12.5\, \mathrm{cMpc}/\rm{h_0}$ since the void center is our simulation center, and the high-resolution void region is a box of length 25$\, \mathrm{cMpc}/\rm{h_0}$.

However, voids are highly aspherical \citep{Ryden.etal.1996,Lavaux.Wandelt.2010,Lavaux.Wandelt.2012, Sutter.etal.2014, Sutter.etal.2015}. Studies have shown that treating voids as spheres can introduce significant bias in terms of constraining neutrino mass using the void size function (VSF) \citep[see e.g.][]{Kreisch.etal.2022}. To account for this, we employ the Hierarchical Equal Area isoLatitude Pixelation (HEALPix)\footnote{http://healpix.sourceforge.net} framework \citep{Gorski.etal.2005, Zonca.etal.2019} and its Python bindings, \texttt{healpy}, to divide our spherical void into 12 regions with equal surface area. For each of these 12 solid angles, we compute the density profile separately.

We find that the overall trend of the density profile remains consistent across the different solid angle bins: very low density at the center, followed by a rise and overshoot of the mean density of the universe, eventually plateauing at larger radii. The variation occurs in the location where the density profile overshoots the mean density, reflecting the aspherical nature of the void. Some directions reach the mean density faster than others, with the minimum radius being $\sim40\, \mathrm{cMpc}/\rm{h_0}$, the maximum radius $\sim120\ \rm cMpc/\rm{h_0}$, and the mean radius $\sim60\ \rm cMpc/\rm{h_0}$, as seen by the 12 grey lines in the background of Figure~\ref{fig:void_density}.

This self-similar shape of void density is driven by the universal properties of evolving spherical voids. Voids expand\footnote{However, this may not be true for the very small voids (with radii of a few Mpc) – they shrink, as the void-in-cloud mechanism describes \citep{Sheth.van.de.weygaert.2004}.}, in contrast to overdense regions, which collapse. The early evolution of such system is well described by spherical expansion \citep[analogous to the spherical collapse model;][]{Peebles.1980}. As they expand, the density within decreases continuously. Outward expansion makes voids evolve towards a spherical geometry. With a minimum density near the void center, and density increasing gradually as one moves outwards, the density deficit of the void decreases as a function of radius. The outward-directed peculiar acceleration is directly proportional to the integrated density deficit, so decreases with radius: inner shells are propelled outwards at a higher rate, so that the interior layers of the void move outwards more rapidly. The inner matter starts to catch up with the outer shells, leading to a steepening of the density profile in the outer realms. Meanwhile, over a growing area of the void interior, the density distribution is rapidly flattening. This is a direct consequence of the outward expansion of the inner void layers; the flat part of the density profile in the immediate vicinity of the dip gets inflated along with the void expansion. At this stage, the void reaches "shell-crossing", and the initial spherical expansion model breaks down and non-linear effects kick in. Shell-crossing occurs at a linearized underdensity $\delta^{L} = -2.8$, or, a physical underdensity of $\delta = -0.81$. Later, voids continue to expand slowly in a self-similar fashion \citep{Fillmore.Goldreich.1984}, with comoving radius $R\propto a^{1/3}$ in an Einstein de-Sitter Universe, significantly slower than the expansion before shell-crossing. \citet{Sheth.van.de.weygaert.2004} showed that whatever the initial void density configuration is, the evolution of a void is always towards a spherical top-hat. This is in stark contrast to how overdensities evolve.

\subsection{Zooming into the void} \label{sec:void_zoom}
\subsubsection{Fiducial void simulation}\label{sec:void_fiducial}

\begin{figure*}
    \centering
    \includegraphics[width=\textwidth]{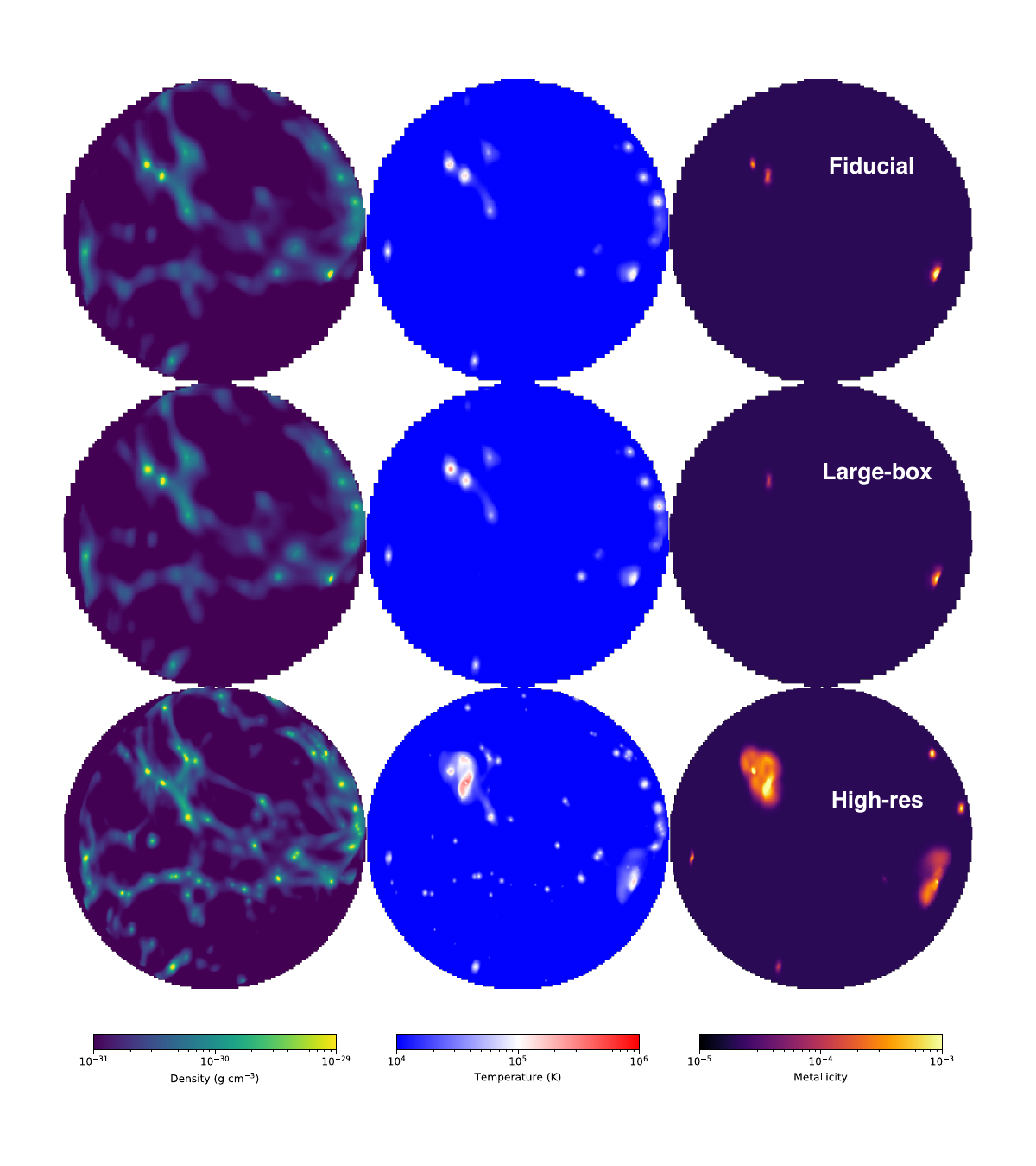}
    \vspace{-20pt}
    \caption{Gas density (left), gas temperature (middle), and gas metallicity (right) for our fiducial 25$\, \mathrm{cMpc}/\rm{h_0}$ sphere void (top), the central 25$\, \mathrm{cMpc}/\rm{h_0}$ sphere for the 50$\, \mathrm{cMpc}/\rm{h_0}$ large box void (middle), and high-resolution 25$\, \mathrm{cMpc}/\rm{h_0}$ sphere void (bottom), all projected along the x-axis.}
    \label{fig:rho_T_Z_3voids}
\end{figure*}

After obtaining the center of our void region, we generate new initial conditions using \texttt{MUSIC}. This time, we define a zoom region of extent 25 cMpc/$h_0$ at $z = 124$ centered on the void coordinates. Outside the zoom region, we progressively degrade the resolution from $\ell = 13$ to $\ell = 9$. Inside the zoom region, we start from $\ell = 13$ and refine the mesh adaptively according to some adopted Adaptive Mesh Refinement (AMR) criterion \citep{Teyssier.2002}. We set the maximum resolution to $\ell = 19$ for this fiducial void dark matter only simulation.

We then incorporate baryons into the simulation. The mass resolution of dark matter particles for the fiducial void dark matter only simulation is $m_{\rm dm, min} \approx 2.3\times 10^8 M_\odot$, calculated using $m_{\rm dm, min} = \Omega_{\rm m, 0} \, \rho_{\rm crit, 0} \, (L/2^\ell)^3$. The corresponding mass resolution of baryons is $m_{\rm b, min} = f_{\rm b} m_{\rm dm, min} \approx 4.6\times 10^7 M_\odot$ adopting $f_{\rm b} = \Omega_{\rm b, 0} / \Omega_{\rm m, 0} \approx 0.2$ as the baryon fraction in the fiducial void hydrodynamical simulation. 

\subsubsection{Larger void simulation}\label{sec:void_larger}

In Fig.~\ref{fig:void_density}, we have found that our fiducial void simulation, with a diameter of 25$\, \mathrm{cMpc}/\rm{h_0}$, is too small – the actual void has an average radius of approximately 60$\, \mathrm{cMpc}/\rm{h_0}$ (or, a diameter of  120$\, \mathrm{cMpc}/\rm{h_0}$), nearly five times larger. Since massive galaxies tend to accumulate at the edge of the void \citep[see e.g.][]{Gottlober.etal.2003}, excluding these edge galaxies in the zoomed region means we are not accurately capturing the physics of star formation and the feedback of these edge galaxies on the void properties within.

To address this issue, we run an additional simulation with a zoom region that covers 8 times more volume. We use \texttt{MUSIC} to generate new initial conditions, centering the zoom region at the same void coordinates with $\ell = 13$, but increasing the extent of the zoom region from 25$\, \mathrm{cMpc}/\rm{h_0}$ at $z = 124$ to 50$\, \mathrm{cMpc}/\rm{h_0}$. We find that the simulation results are visually identical to the fiducial void simulations in terms of the halo mass function, the distribution of star particles, gas metallicity, and gas density, indicating that the massive galaxies at the edge of the larger void do not have an immediate effect on the physics within the central 25$\, \mathrm{cMpc}/\rm{h_0}$ region of the void. Fig~\ref{fig:HMF} shows the nearly identical cumulative halo mass function for fiducial void simulations (dark matter only and hydro) and for the larger box void simulations.

\begin{figure*}
    \centering
    \includegraphics[width=\textwidth]{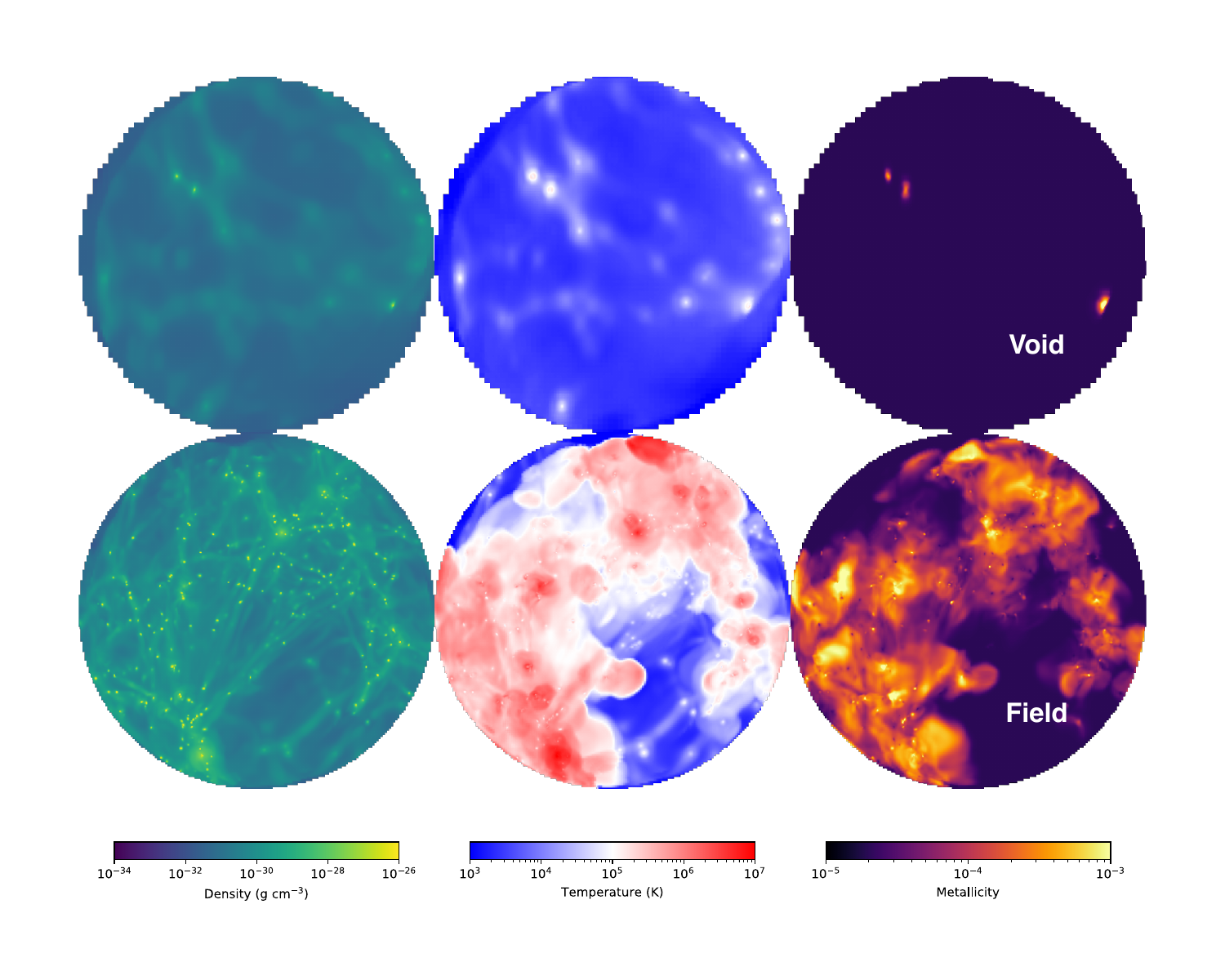}
    \vspace{-20pt}
    \caption{Gas density (left), gas temperature (middle), and gas metallicity (right) for our 25$\, \mathrm{cMpc}/\rm{h_0}$ sphere in the void (top) and 25$\, \mathrm{cMpc}/\rm{h_0}$ sphere in the field (bottom), both projected along the x-axis.}
    \label{fig:rho_T_Z_camels_void}
\end{figure*}

\subsubsection{High-resolution void simulation}\label{sec:high_red_void}

Since the galaxies on the rim of the 120$\, \mathrm{cMpc}/\rm{h_0}$ size void do not affect the physics of the galaxies in the central 25$\, \mathrm{cMpc}/\rm{h_0}$ region, we now focus solely on this central region to explore galaxy formation and galactic winds. We then compare these properties to those in the field simulation to understand how void environments shape galaxy evolution. 

Furthermore, \citet{Wang.Pisan.2024} demonstrates that cosmological information can be extracted from a galaxy within a void. This finding suggests that galaxy properties are influenced not only by their local surroundings but also by the broader cosmic environment. Given that their analysis is based on simulations of 25$\, \mathrm{cMpc}/\rm{h_0}$ voids, this supports the idea that 25$\, \mathrm{cMpc}/\rm{h_0}$ is a reasonable scale for studying the impact of void environments on galaxy evolution.

To further study the detailed physics in void, we add one additional level of resolution ($\ell = 14$) to our target region with a diameter of 25$\, \mathrm{cMpc}/\rm{h_0}$ in order to study the effect of numerical resolution on the properties of void galaxies.

Fig.~\ref{fig:rho_T_Z_3voids} compares the gas density, gas temperature, and gas metallicity between the three void simulations. The larger box void simulation yields gas properties identical to those of the fiducial void simulation, while the high-resolution void simulation produces many more small mass galaxies. It also resolves much more detailed structure in the gas and provides a clearer depiction of outflows. The extent of outflows seen in the gas temperature and gas metallicity panels for the few large galaxies in the void is much more extended in high-resolution void simulation than the fiducial void simulation.


\subsection{Field simulation for comparison} \label{sec:camels_hydro}

To compare the void simulation with a field environment, we use \texttt{MUSIC} to generate new initial conditions for a comoving volume $L^3 = (25\rm\, cMpc/h_0)^3$ at $z = 124$ without zoom at a resolution of $\ell = 8$. This field simulation is conditioned with zero overdensity, and the mass resolution of dark matter particles is $\sim 1.2\times 10^8 M_\odot$, a factor of two more resolved than the fiducial void simulation but an order of magnitude less resolved than the high-resolution void simulation. 

Figure~\ref{fig:rho_T_Z_camels_void} shows a comparison of gas density, gas temperature, and gas metallicity between the 25$\, \mathrm{cMpc}/h_0$ sphere in the field and void simulations.  From this simple visual inspection, the lack of structures, outflows, and low gas temperature in our void is quite striking. There are a few massive galaxy clusters in the field. As can be seen in the gas density panel, the clusters are located in the junction of structures in a high density region. 

\begin{figure*}
    \centering \includegraphics[width=\textwidth]{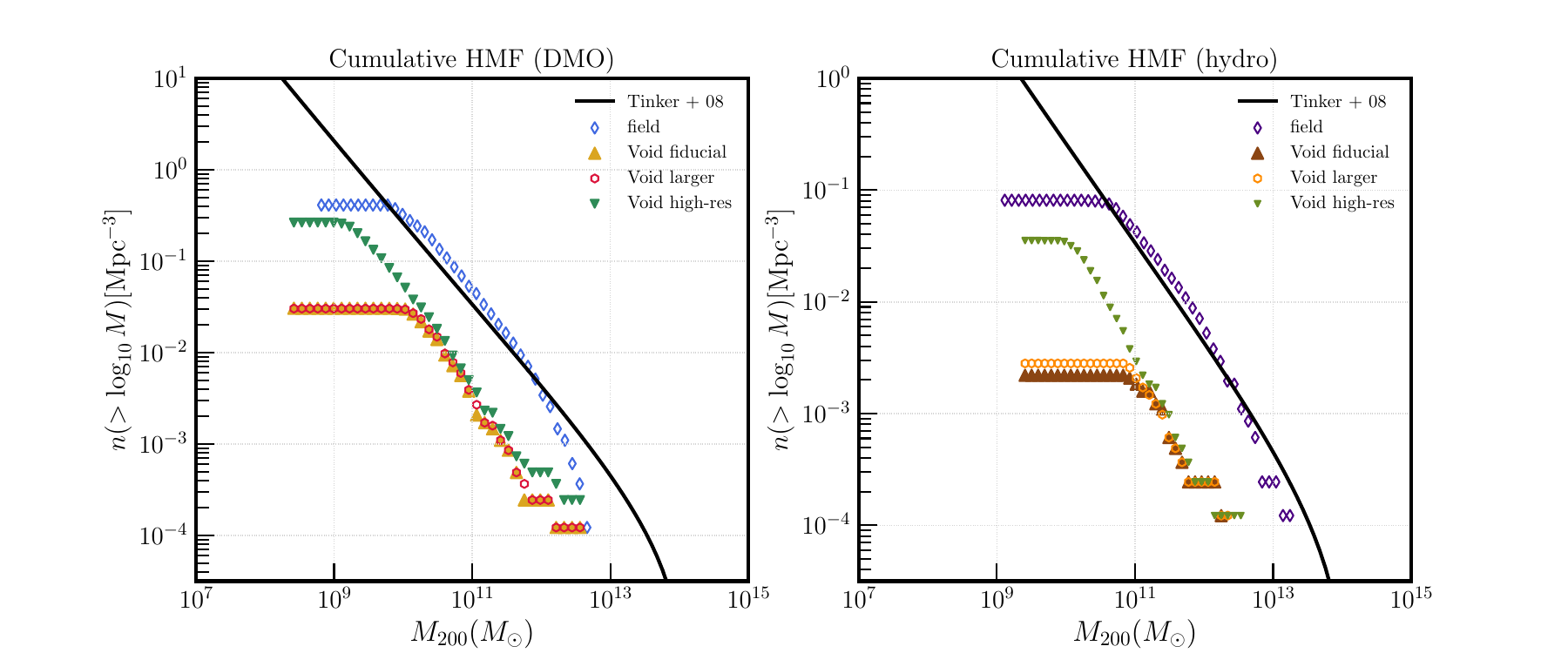}
    \caption{Cumulative halo mass function for all dark matter-only simulations (left) and hydrodynamical simulations (right), using our computed halo mass $M_{200}$. For comparison, we overplot the results from \citet{Tinker.etal.2008}. The halo mass function for the larger-box void and fiducial void simulations are identical for both dark matter only and hydrodynamical runs, indicating that galaxies located at the edge of the true $120 \, \mathrm{cMpc}/\rm{h_0}$ void do not influence the physics of the inner $25 \, \mathrm{cMpc}/\rm{h_0}$ region. However, increasing the resolution by one level has a substantial impact on the halo mass function, resulting in nearly an order of magnitude more small halos (with $M_{\rm h} \sim 10^9 \, M_\odot$ for dark matter only and $M_{\rm h} \sim 10^{9.5} \, M_\odot$ for hydro) being resolved. Since most galaxies in the void are dwarfs, the high-resolution run includes significantly more galaxies (30 compared to 6) than the fiducial run within the same box size of $25 \, \mathrm{cMpc}/\rm{h_0}$.
} 
    \label{fig:HMF}
\end{figure*}

\subsection{Post-process halo finder (PHEW)} \label{sec:clumpfind}
To construct the halo mass function, we need accurate positions of all the halos in the simulation. The built-in clump finder algorithm, PHEW \citep[Parallel HiErarchical Watershed,][]{Bleuler.etal.2015}, in \texttt{RAMSES} finds density peaks and associated regions in the 3D density field by performing watershed segmentation. Then it merges substructures based on the saddle point topology in two steps: first, merge density fluctuations (noise);  second, merge finest substructure hierarchically into large, connected regions. 

However, the PHEW algorithm initially identifies dense regions by tracing gas particles, effectively locating clumps of gas within the simulation. These dense gas clumps serve as markers, or ``seeds'', for the formation of supermassive black holes, which is crucial for calculating feedback effects that impact galaxy formation and evolution. Therefore, accurately identifying these dense clumps during the simulation is essential for correctly modeling feedback mechanisms. However, because this clump-finding process is based on gas density, it does not necessarily track the underlying dark matter distribution accurately. For a robust identification of halos – gravitationally bound structures dominated by dark matter – the algorithm should ideally locate dense clumps of dark matter rather than gas. To address this, a second iteration of clump-finding is conducted post-process on the final snapshot, specifically targeting dark matter density. This additional step allows for a more accurate mapping of dark matter halos, which is necessary for studying the structure and distribution of matter in the Universe.

\section{Halo statistics}\label{sec:halo_stats}

Our goal is to derive a theoretical form of the stellar mass function extending from the largest halo in the void and field simulation down to the smallest halo capable of hosting a galaxy (i.e., the filtering mass). Using this model, combined with wind modeling that calculates wind volume from supernova explosion based on galaxy's stellar mass  (Section~\ref{sec:wind_modeling}), we aim to predict the theoretical volume filling factor of void and field galaxies, assuming infinite resolution in the simulation.

To achieve this, we need an accurate model for the stellar mass function. We start with an analytical halo mass function, fit it to the simulated halo mass function, and then apply a fixed theoretical halo occupation fraction and stellar mass--halo mass relation to derive the stellar mass function. We choose to fit for the halo mass function as it is a relatively well-established quantity in terms of the shape, slope, and physical origin. Consequently, the first and most crucial step is to obtain an accurate theoretical fit for the halo mass function that is consistent with both the field simulation and our high-resolution void simulation.

\subsection{Halo mass function}\label{sec:HMF}

To calculate halo mass, we expand concentric shells from the halo center identified by PHEW, defining the halo boundary $r_{200}$ as the radius at which the total enclosed density drops to 200 times the critical density of the universe at $z=0$. The corresponding enclosed mass, $M_{200}$, is taken as the halo mass. Within $r_{200}$, the sum of star particle masses defines the galaxy stellar mass, $M_\star$.

To ensure that massive particles outside the high-resolution region do not contaminate halo mass statistics, we select a true "void" region free from massive particles. Since the original high-resolution void box of length $25 \, \mathrm{cMpc}/\rm{h_0}$ expanded and its center shifted, we test several spheres with diameters of $25 \, \mathrm{cMpc}/\rm{h_0}$ near the box center to assess contamination by massive particles. By slightly adjusting the sphere in the $z$-direction, we obtain a region entirely free of massive particles, allowing for accurate computation of halo statistics within this spherical region.

In \texttt{RAMSES}, PHEW also returns an alternative halo mass definition, defined as the enclosed mass within an isodensity contour of 80 times the critical density of the Universe. We compare halo masses derived using both definitions and find good agreement. For consistency with subsequent calculations involving halo stellar masses, and since the PHEW method does not return stellar masses, we use the spherical overdensity definition for $M_{200}$  as the halo mass proxy throughout our analysis.

Fig.~\ref{fig:HMF} presents the cumulative halo mass functions for all 8 simulations. Comparing the dark matter-only runs to their hydrodynamical counterparts (e.g., fiducial void dark matter only and fiducial void hydro), we observe baryonic effects that disrupt low-mass halos, resulting in a flattening of the halo mass function at the low-mass end relative to pure dark  matter only runs. Furthermore, comparing field dark matter only with fiducial void dark matter only (or field hydro with fiducial void hydro), we note a significant reduction in halo counts across all mass scales. 

The low-mass end flattening of each halo mass function curve indicates the mass resolution limits for each simulation set. Specifically, the high resolution void simulation achieves the highest mass resolution of $2.9 \times 10^7 \, M_\odot$, followed by the field simulation with a mass resolution of $1.2 \times 10^8 \, M_\odot$, and finally fiducial void simulation with a mass resolution of $2.3 \times 10^8 \, M_\odot$.

\subsubsection{Halo mass function in the field}\label{sec:CAMELS_HMF}

\begin{figure*}
    \centering
    \includegraphics[width=\textwidth]{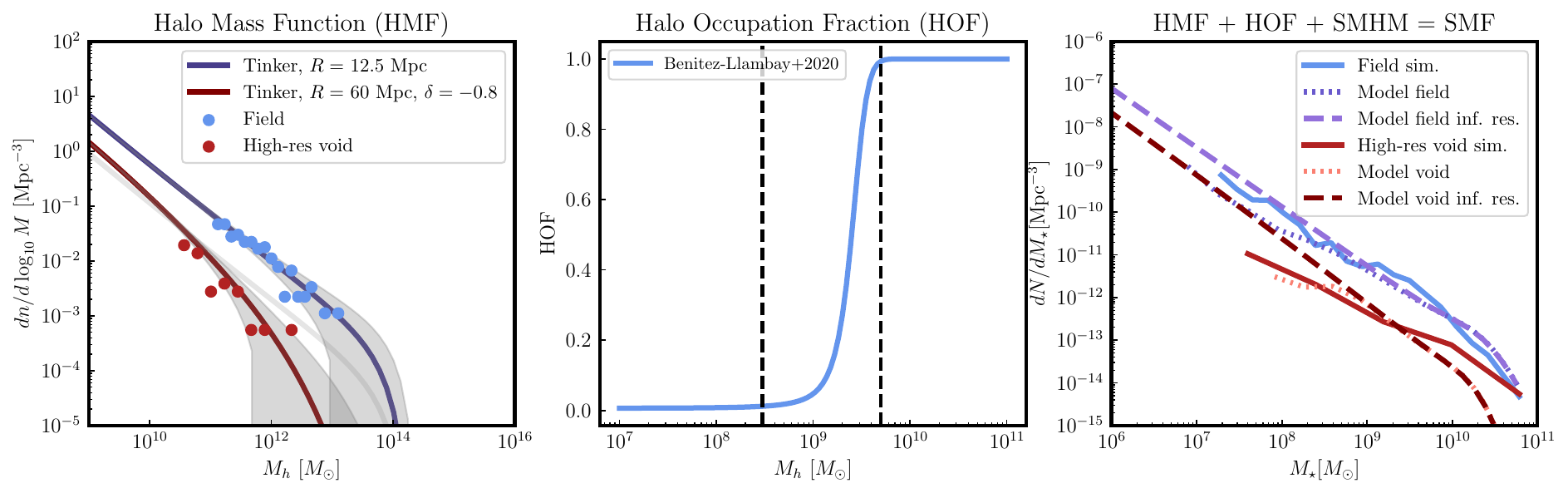}
    \caption{\emph{Left:} The halo mass function for the field simulation data (blue points) and the high-resolution void (red points) is shown alongside theoretical fits: \citet{Tinker.etal.2008} (purple line), rescaled Tinker by $1+\delta_{\rm void}$ to account for void expansion (light grey), and \citet{Furlanetto.Piran.2006} (red line). Poisson error bars on the theoretical fits (shaded gray area) are described in detail in Section~\ref{sec:CAMELS_HMF}. \emph{Middle:} The halo occupation function derived from \citet{Benitez-Llambay.Frenk.2020} is displayed, which we approximate using a sigmoid function (Eqn~\ref{eqn:hof}). This function is constrained by two boundary conditions: an occupation fraction of 0\% below the filtering mass $M_{\rm h} = 3\times 10^8 M_\odot$ and 100\% above $5\times 10^9 M_\odot$. \emph{Right:} The stellar mass function for the field simulation is represented by the blue solid line, while our model predictions—combining the halo mass function, halo occupation fraction, and stellar mass--halo mass relation—are shown as the purple dotted line. For the infinite-resolution case, the corresponding predictions are shown as the purple dotted-dashed line. Similar results are presented for the high-resolution void, with analogous lines. Notably, the halo mass function derived from our model agrees well with the simulation data for both the field and the high-resolution void in the resolution-limited case (blue vs. purple dotted line, red vs. pink dotted line).} 
    \label{fig:HMF_HOF_SMHM}
\end{figure*}

The halo mass function in the field (e.g., CAMELS environment) was initially derived analytically by \citet{Press.Schechter.1974} from spherical collapse theory for virialized objects in a hierarchical density field $\delta(\vec x, R)$. They assumed that objects will collapse on some scale once the smoothed density contrast on this scale exceeds a threshold value, and that the nonlinearities introduced by these virialized objects do not affect the collapse of overdense regions on larger scales. This assumption is only approximately true, but in voids, where the additional large-scale power generated by nonlinearities of virialized objects is much smaller than the primordial fluctuations on these scales, it is a fairly valid approximation.

In mathematical terms, a halo of mass $M_{\rm h}$ forms whenever the smoothed density field first exceeds a linearly-extrapolated density threshold $\delta_{\rm c}^{\rm L}$, determined by the physics of spherical halo collapse \citep[$\delta_{\rm c}^{\rm L} = 1.686$ for objects collapsing at the present day;][]{Peebles.1980}. Furthermore, for objects whose smoothed density field has already surpassed the collapse threshold before $z = 0$, mass growth continues, so their mass at $z = 0$ is greater than that at the time of collapse. Consequently, to count the number density of halos at the present day within a narrow mass bin, we essentially count the number density of virialized objects with mass $> M$ at collapse. To obtain the differential halo mass function, we take the derivative of this cumulative number density.

However, the fraction of all mass in virialized objects integrates to $1/2$ under this assumption, implying that only half of the mass density of the universe is contained in virialized objects. To resolve this, \citet{Press.Schechter.1974} introduced a fudge factor of 2 to account for underdense regions collapsing onto overdense regions that were not included in the original calculation. This is referred to as the \emph{cloud-in-cloud problem}, as it miscounts the number of low-mass clumps in the literature. A limitation of the Press-Schechter approach is that it does not account for the fact that, at a given smoothing scale and location $\vec{x}$, $\delta(R)$ may be less than $\delta_{\rm c}$ but could exceed $\delta_{\rm c}$ on a larger scale $R' > R$. Consequently, this larger volume should collapse to form a virialized object, overwhelming the more diffuse patches within it.

To address this issue, it is essential to calculate the largest smoothing scale, $R$, at which the density field first exceeds the critical threshold, $\delta_{\rm c}$. \citet{Bond.etal.1991} introduced an elegant solution to this problem by linking the sequence of density values, $\delta(R_{\rm i})$, obtained through successive increases in the smoothing scale, to a continuous \emph{trajectory} of $\delta(R)$. For a sharp Fourier-space filter applied to the density field, these trajectories exhibit stochastic behavior analogous to a Brownian random walk \citep{Chandrasekhar.1943}. However, for more general smoothing filters, such as Gaussian or top-hat filters, deriving analytic solutions becomes significantly more challenging. In such cases, approximate solutions can be obtained by numerically generating a large number of trajectories using a Langevin equation to simulate the random motion of the density field \citep[see, e.g.,][]{Chandrasekhar.1943}. The objective of this approach is to determine the largest smoothing scale, $R$, at which a trajectory first crosses the barrier defined by $\delta_{\rm c}$. This scale provides critical insights into the hierarchical growth of structures in the universe and forms the basis for understanding halo formation in the excursion set formalism.

By examining the Brownian trajectory of the smoothed density field, $\delta$, across varying smoothing scales $R$, \citet{Bond.etal.1991} demonstrated that the fraction of trajectories absorbed by the barrier $\delta_{\rm c}$ at scales larger than $R$ corresponds to the fraction of mass in objects with masses greater than $M$. This framework addresses a critical limitation in the original Press-Schechter model by accounting for objects that fail to exceed $\delta_{\rm c}$ at scale $R$ but cross the threshold at a larger scale $R' > R$. These objects are thus correctly assigned to collapsed halos with mass $> M$. In contrast, the original Press-Schechter formalism assumed that the fraction of mass in halos of mass $> M$ is determined solely by the fraction of mass above $\delta_{\rm c}$ at scale $R$. This assumption neglects contributions from trajectories that cross the barrier at larger scales, leading to an underestimation of the halo abundance by a factor of 2. The incorporation of these additional contributions forms the basis of the \emph{extended Press-Schechter} or \emph{excursion set} formalism. This refined approach has proven effective in capturing the dynamics of structure formation. In particular, it reproduces halo merger rates and formation times with much greater accuracy compared to the original model \citep[e.g.,][]{Lacey.Cole.1993}, offering a more robust theoretical framework for understanding hierarchical structure growth.

Mathematically, it is similar to solving the diffusion problem without drift or a one-dimensional heat equation describing heat flow in a long, thin, semi-infinite bar with boundary condition at $\delta = \delta_{\rm c}$: in the $(\delta, \sigma^2)$ plane where $\sigma^2$ represents the variance of the density field smoothed on a scale $M_{\rm h}$, trajectories begin at $\delta = 0$ and $\sigma^2 = 0$ (i.e., at mean density on infinitely large scales) and then diffuse away from the origin as density modes are added on smaller scales. The problem is to compute the distribution of $\sigma^2$ (or equivalently $M_{\rm h}$) at which these trajectories first cross the linearly extrapolated absorbing barrier $\delta_{\rm c}^{\rm L}$. From this first-crossing distribution, the mean comoving number density of halos with masses $M_{\rm h} \pm dM_{\rm h}/2$ is given by \citep{Press.Schechter.1974, Bond.etal.1991}
\begin{equation}
    n_{\rm h}(M_{\rm h}) = \sqrt{\frac{2}{\pi}} \frac{\bar{\rho}}{ M_{\rm h}^2} \left|\frac{d \ln \sigma}{d \ln M_{\rm h}}\right| \frac{\delta_{\rm c}^{\rm L}}{\sigma} \exp\left(-\frac{\left(\delta_{\rm c}^{\rm L}\right)^2}{2\sigma^2}\right),
    \label{eqn:normal_HMF}
\end{equation}
where $\bar{\rho}$ is the mean matter density, and $\delta_{\rm c}^{\rm L}$ is the critical density for spherical collapse linearly extrapolated to the present day, which we take to be $\delta_{\rm c}^{\rm L} = 1.686$ \citep{Peebles.1980}.

The excursion set formalism has been shown to agree well with results from simulations \citep[e.g.,][]{Efstathiou.etal.1988, Bond.etal.1991, Jenkins.etal.2001, Zheng.etal.2024} and to match observational data to a reasonable extent \citep{Percival.2001}. However, the formalism relies on several simplifying assumptions, which introduce notable limitations \citep[e.g.,][]{Jenkins.etal.2001, Sheth.Tormen.2002, Benson.etal.2005, Li.etal.2007}. One key simplification is the assumption that the critical density threshold, $\delta_{\rm c}$, is independent of both scale and the specific properties of the density field. To address this, \citet{SMT2001} and \citet{Sheth.Tormen.2002} proposed a model incorporating ellipsoidal collapse, rather than spherical collapse, of overdense regions. In this model, the ellipsoidal collapse barrier is higher for low-mass objects, driven by the fact that such objects exhibit greater ellipticity and are more susceptible to tidal disruption. Consequently, these objects must achieve higher densities to resist tidal forces and collapse successfully \citep{SMT2001}. This modification reduces the overprediction of low-mass halos inherent in the constant-barrier excursion set mass function.

Moreover, in the standard excursion set formalism, the density field is smoothed using a sharp $k$-space window function. This choice ensures that transitions between smoothed densities are treated as independent Gaussian random variables, resulting in uncorrelated random walks in the density trajectory. Consequently, the formation histories of halos, as predicted by this formalism, are entirely independent of their local density field, or equivalently, their environments. However, a growing body of evidence from cosmological simulations suggests that halo formation histories are strongly correlated with their large-scale environments \citep{Sheth.Tormen.2004, Gao.etal.2005, Wechsler.etal.2006, Wetzel.etal.2009}. These findings highlight the limitations of neglecting correlations between scales in the standard formalism. To address this issue, the sharp $k$-space window function can be replaced with a smoother window function, such as a Gaussian. This modification introduces correlations into the random walks of density variations, reflecting a more realistic interaction between scales. The inclusion of these correlations has important implications for the predicted halo mass function. Specifically, it leads to a suppression in the number of halos at all mass scales relative to the predictions of the constant-barrier excursion set formalism, with the effect being particularly pronounced for low-mass halos. This refinement not only aligns better with simulation results but also improves the theoretical consistency of the model.

Despite its limitations, the extended Press-Schechter formalism has served as a foundation for developing accurate universal functions to describe halo abundances in simulations \citep{Jenkins.etal.2001, White.2001, Warren.etal.2006}. Building on these early efforts, \citet{Tinker.etal.2008} proposed a functional form that better captures halo abundance in collisionless cosmological simulations. The halo mass function in this formalism is expressed as:
\begin{equation}
    \frac{dn}{dM} = f(\sigma) \frac{\bar{\rho}_m}{M} \frac{d\ln\sigma^{-1}}{dM},
\end{equation}
where $f(\sigma)$ is a universal function that incorporates redshift and cosmology dependence. It is defined as:
\begin{equation}
    f(\sigma) = A \left[ \left(\frac{\sigma}{b}\right)^{-a} + 1 \right] e^{-c/\sigma^2}.
\end{equation}
Here, $\sigma^2$ is the variance of the linear matter density field, given by:
\begin{equation}
    \sigma^2(M) = \frac{1}{(2\pi)^3}\int_0^{+\infty} P(k)W(kR) 4\pi k^2 \, dk,
\end{equation}
where $P(k)$ is the linear matter power spectrum as a function of the wavenumber $k$, $W(kR)$ is the top-hat window function, and $M = 4\pi/3\bar\rho_m R^3$. Since the matter variance $\sigma$ decreases monotonically with increasing smoothing scale, larger halo masses $M$ correspond to smaller values of $\sigma$. The parameters $A$, $a$, $b$, and $c$ are constants calibrated using simulations. Each parameter has a specific role: $A$ sets the overall amplitude of the mass function, $a$ and $b$ control the slope and normalization of the low-mass power-law regime, and $c$ determines the cutoff scale where the halo abundance exponentially declines. This functional form is preferred over the extended Press-Schechter formalism because it is calibrated against simulations, allowing it to capture more detailed physics. Moreover, it accounts for the dependence of halo abundance on cosmology and redshift, making it a more accurate and flexible framework for describing halo mass functions.

In the left panel of Fig.~\ref{fig:HMF_HOF_SMHM}, we present the halo mass function from \citet{Tinker.etal.2008}, conditioned on the field or void density deficit $\delta$ and smoothing scale $R$. These results are compared against the halo mass function data points derived from our simulation. Encouragingly, the simulation data points lie well within the Poisson error bars of the analytical expression, providing strong validation for the theoretical halo mass function. This agreement lends confidence to subsequent calculations, particularly those involving the volume filling factor derived from this halo mass function.

To compute the Poisson error bars for the analytical curves, first, we calculate the expected number of halos $dn$ in each mass bin by multiplying the differential halo mass function $dn/d\log_{10}M$ by the width of the mass bin, $dM$, and the volume of the simulation, $V = (4/3)\pi(L/2)^3$, where $L = 25 \, \mathrm{Mpc}$. The corresponding Poisson uncertainty is then given by $\sqrt{dn}$. Finally, the error bars are expressed as fractional uncertainties, computed as $(dn/d\log_{10} M) / \sqrt{dn}$.

\subsubsection{Halo mass function in the void}\label{sec:void_HMF}
Fig.~\ref{fig:HMF} highlights the significant suppression of halos across all mass scales in voids compared to the field. In particular, the most massive halos present in voids have a mass of approximately $M_{\rm h} \sim 10^{12.5} M_\odot$, which is comparable to the mass of the Milky Way \citep{Cautun.etal.2020, Shen.etal.2022}. This finding underscores the fact that only small halos, with masses lower than that of the Milky Way, can form in such low-density void regions due to the lack of sufficient matter to facilitate the growth of more massive structures. Moreover, it becomes evident that standard halo mass function models, such as those proposed by \citet{Schechter.1976}, \citet{SMT2001}, or \citet{Tinker.etal.2008}, fail to accurately describe the halo abundance in voids, as the overall matter density is significantly less than the field. Consequently, alternative approaches that account for the unique dynamics of void regions are necessary to accurately model halo formation and abundance in these settings.

To study how halos collapse and the number density of virialized halos in voids, we condition the Press-Schechter formalism for the halo mass function in the field (Eq.~\ref{eqn:normal_HMF}) on void properties, $M_{\rm v}, \delta_{\rm v}^{\rm L}$, where $M_{\rm v}$ is the total mass within the void, and $\delta_{\rm v}^{\rm L}$ represents the linearized void underdensity. This conditional halo mass function (or, equivalently, extended Press-Schechter) can be written in comoving units as  \citep{Bond.etal.1991, Lacey.Cole.1993, Furlanetto.Piran.2006}:
\begin{align}
    n_{\rm h}(M_{\rm h} | \delta_{\rm v}^{\rm L}, M_{\rm v}) &= \sqrt{\frac{2}{\pi}} \frac{\bar{\rho}}{M_{\rm h}^2} \left|\frac{d \ln \sigma}{d \ln M_{\rm h}}\right| \frac{\sigma^2 (\delta_{\rm c}^{\rm L} - \delta_{\rm v}^{\rm L})}{(\sigma^2 - \sigma_{\rm v}^2)^{3/2}} \nonumber \\
    &\quad \times \exp\left[-\frac{(\delta_{\rm c}^{\rm L} - \delta_{\rm v}^{\rm L})^2}{2(\sigma^2 - \sigma_{\rm v}^2)}\right],
    \label{eqn:void_HMF}
\end{align}
where $\sigma_{\rm v} \equiv \sigma(M_{\rm v})$ is \textbf{the square root of} the variance of the density field smoothed on the void mass scale, and $\delta_{\rm c}^{\rm L}$, $\bar{\rho}$, and $\sigma$ are defined the same as Eq.~\ref{eqn:normal_HMF}. Since voids encompass scales larger than the halos they contain, it follows that $\sigma > \sigma_{\rm v}$, with $\sigma$ decreasing as the smoothing scale increases. This relationship ensures that the variance accurately reflects the hierarchical structure of the universe, where larger structures such as voids exhibit lower density fluctuations than the smaller halos within them.

To incorporate the effects of the void environment on the halo mass function, we compute the ratio of the void halo mass function to the field halo mass function as a function of halo mass, $M_{\rm h}$. This ratio is then applied to the \citet{Tinker.etal.2008} halo mass function, which provides a more accurate parametrization of the field halo mass function than Press-Schechter as it is calibrated against simulations.  We use the provided parametrization of the Tinker halo mass function in \texttt{Colossus} \footnote{https://bdiemer.bitbucket.io/colossus/}\citep{Diemer2018}.

The ratio between the void halo mass function, $n_{\rm h}(M_{\rm h} | \delta_{\rm v}^{\rm L}, M_{\rm v})$ (Eq.~\ref{eqn:void_HMF}), and the standard field halo mass function, $n_{\rm h}(M_{\rm h})$ (Eq.~\ref{eqn:normal_HMF}), is given by:
\begin{equation}\label{eqn:ratio}
    \frac{n_{\rm h}(M_{\rm h} | \delta_{\rm v}^L, M_{\rm v})}{n_{\rm h}(M_{\rm h})} = \frac{\sigma^3 (\delta_{\rm c}^{\rm L} - \delta_{\rm v}^{\rm L})}{(\sigma^2 - \sigma_{\rm v}^2)^{3/2} \delta_{\rm c}^{\rm L}} 
    \exp\left[-\frac{(\delta_{\rm c}^{\rm L} - \delta_{\rm v}^{\rm L})^2}{2(\sigma^2 - \sigma_{\rm v}^2)} + \frac{(\delta_{\rm c}^{\rm L})^2}{2\sigma^2}\right],
\end{equation}
where all parameters are defined as before. This ratio provides a framework for understanding how void conditions suppress the abundance of halos relative to the field and offers a means to adapt standard halo mass functions to accurately reflect the distinct dynamics of underdense regions.

To calculate the variance of the density field at the initial collapse scale of the void, $\sigma_{\rm v}$, we use Pylians\footnote{\url{https://pylians3.readthedocs.io/en/master/mass_function.html}} \citep{Pylians} and its "mass function library". For our specific void, its physical underdensity is $\delta = \bar{\rho}/\rho_{\rm crit} - 1 = 0.2 - 1 = -0.8$ (Section~\ref{sec:void_prop}). According to Figure 1 in \citet{Furlanetto.Piran.2006}, for our physical void deficit of $\delta = -0.8$, the corresponding linearly extrapolated void underdensity to the present day is approximately $\delta_{\rm v}^{\rm L} \approx -3$. Since void expands and the void radius at the present day is approximately 60 Mpc (Section~\ref{sec:void_prop}), the initial collapse radius is scaled down by a factor of $(1 + \delta_{\rm v}^{\rm L})^{-1/3} \approx 1.7$, resulting in an initial radius of $60/1.7 \approx 35$ Mpc. The variance $\sigma_{\rm v}$ is then evaluated at this radius using mass function library in Pylians with the linear power spectrum at $z=0$ derived from the Code for Anisotropies in the Microwave Background \citep[CAMB;][]{Lewis.etal.2011}. 

Using these values, we compute the ratio 
\(\frac{n_{\rm h}(M_{\rm h} | \delta_{\rm v}^{\rm L}, M_{\rm v})}{n_{\rm h}(M_{\rm h})}\) 
at 100 logarithmically spaced halo mass values ranging from \(10^9 M_\odot\) to \(10^{14} M_\odot\). This calculation provides a detailed characterization of how the halo mass function is modified under void conditions. However, this ratio is expressed in comoving units. Voids expand over cosmic time due to the mass deficit at their centers. Consequently, the physical number density of halos within the void is diluted by the volume increase, which scales as \(1+\delta\), where \(\delta\) represents the physical underdensity of the void.

Figure~\ref{fig:HMF_HOF_SMHM} illustrates the binned halo mass function in the simulations alongside our analytical fits. The left panel shows the binned halo mass function in the field simulation (blue points), which is fitted using the Tinker mass function (purple line). The void simulation results are represented by maroon points and are fitted with a rescaled version of the Tinker mass function (grey line) using a factor of \(1+\delta\) to account for the void's expansion. Additionally, the maroon line includes further modifications to incorporate variations across different halo mass scales, reflecting the halo mass function conditioned on the void environment. 

To summarize, the analytical void halo mass function is derived through the following three steps:
\begin{enumerate}
    \item We begin with the halo mass function from \citet{Tinker.etal.2008}, which is calibrated against simulations and provides a more accurate parametrization of the field halo mass function.
    \item Next, we apply Eq.~\ref{eqn:ratio} to modify the Tinker halo mass function, incorporating the constraint imposed by the void environment.
    \item Finally, we introduce an additional factor of \(1+\delta_{\rm void}\) to account for the dilution of the void halo mass function due to the expansion of voids over cosmic time. The final result is the maroon line in the left panel of Figure~\ref{fig:HMF_HOF_SMHM}.
\end{enumerate}

Notably, the void halo mass function exhibits a relative surplus of low-mass halos compared to the rescaled Tinker mass function at the low-mass end (\(M_{\rm h} < 10^{10.2} M_\odot\)) and a significant deficit at the high-mass end (\(M_{\rm h} > 10^{10.2} M_\odot\)). These trends highlight the distinct dynamics of halo formation in underdense environments compared to the field.

The agreement between the high-resolution void simulation data and the fitted void halo mass function derived purely from theory is remarkable. Similarly, the binned halo mass function from the field simulation aligns closely with the Tinker mass function. This consistency validates the use of the void halo mass function for underdense regions and the Tinker mass function for the field in subsequent theoretical calculations. In particular, it lends credibility to their application in determining the theoretical volume filling factor in later analyses. 

Moreover, the void halo mass function has been explored as a probe for cosmological parameters, including neutrino mass, due to the sensitivity of underdense regions to their effects. While previous studies suggested its potential, \citet{Bayer.etal.2024} emphasized the role of void shapes, showing via VIDE that the void halo mass function alone offers no additional constraints beyond the standard halo mass function. However, they introduced the Voronoi halo mass function, which classifies halos by their local ``emptiness'' and shows promise in improving neutrino mass constraints. These findings underscore the need for precise void mass function modeling in cosmological studies.

\subsection{Halo occupation fraction (HOF)}\label{sec:SMHM}

To quantify the fraction of halos in each mass bin that host galaxies, we employ the Halo Occupation Fraction (HOF) to connect the total number of halos to the subset that are luminous. Following the formalism outlined by \citet{Benitez-Llambay.Frenk.2020}, we adopt their conclusion that, at the present day, halos with masses below $M_{\rm h} < 3 \times 10^8 M_\odot$ are unable to form stars, while those with masses exceeding $M_{\rm h} > 5 \times 10^9 M_\odot$ host galaxies with 100\% efficiency. These cutoff values are supported by independent studies. For instance, \citet{Okamoto.etal.2008} found that the characteristic mass at which halos lose, on average, half of their baryons to photo-heating is approximately $5 \times 10^9 M_\odot$. Similarly, \citet{Sawala.etal.2016}, using hydrodynamical simulations that incorporate sub-grid physics, suggest that the threshold mass above which all halos are occupied is close to $10^{10} M_\odot$ at the present day. These findings collectively reinforce the adopted halo occupation fraction framework and its associated mass cutoffs.

The inability of halos with $M_{\rm h} < 3 \times 10^8 M_\odot$ to host luminous galaxies is due to their shallow potential wells, which fail to retain gas against the pressure exerted by the external UV radiation field. In contrast, halos with $M_{\rm h} > 5 \times 10^9 M_\odot$ possess sufficiently deep potential wells such that the gravitational acceleration they generate requires an amount of gas far exceeding the universal baryon fraction, $\Omega_{\rm b}/\Omega_{\rm m}$, to remain in hydrostatic equilibrium. Consequently, gas in these halos cannot remain stable and will collapse to form stars, ensuring that every such halo hosts a luminous galaxy \citep{Benitez-Llambay.Frenk.2020}.

For halos with masses in the range $3 \times 10^8 M_\odot < M_{\rm 200} < 5 \times 10^9 M_\odot$, whether or not the halo hosts a luminous galaxy depends sensitively on its mass assembly history. If a halo within this mass range crosses the critical mass threshold required for gas to collapse at any point in its history, it can host a luminous galaxy at $z = 0$; otherwise, it remains dark. The gradual transition of the halo occupation fraction from 0 to 1 in this mass range reflects the stochastic nature of halo mass growth, where variations in accretion histories and environmental interactions determine whether a halo reaches the critical mass for star formation \citep{Hoeft.etal.2006, Benitez-Llambay.etal.2015, Sawala.etal.2016, Fitts.etal.2017, Munshi.etal.2019, Wheeler.etal.2019}.

We approximate the halo occupation fraction using the two boundary values and fit it with a sigmoid function:
\begin{equation}
    S(M_{\rm h}) = \frac{1}{1 + \exp(-k(M_{\rm h} - M_0))},
\end{equation}\label{eqn:hof}
where the parameters are set to $k = 2 \times 10^{-9}$ and $M_0 = 2.5 \times 10^9 M_\odot$. The resulting form of the halo occupation fraction is illustrated in the middle panel of Fig.~\ref{fig:HMF_HOF_SMHM}, showing the two limiting cases. We adopt the sigmoid function to approximate the halo occupation fraction due to its smooth transition between the boundary values of 0 and 1, which reflects the gradual change in halo occupation. The purpose of this approximation is to provide a reasonable and practical model for the halo occupation fraction, rather than the most accurate form derived from detailed observations or simulations. As discussed in the following paragraph, we consider this approximation sufficient for our analysis and employ it consistently in all subsequent calculations.

We emphasize that while the adopted form of the halo occupation fraction is consistent with high-resolution cosmological hydrodynamical simulations incorporating reionization in the redshift range $6 < z_{\rm rei} < 10$ \citep{Planck.2018}, its precise functional form remains a subject of ongoing debate. For instance, \citet{Nadler.etal.2020} employed a statistical framework to parameterize both the halo occupation fraction and the stellar mass--halo mass relation, fitting these parameters to the observed Milky Way satellite population identified in the Dark Energy Survey (DES). Their results suggest that the faintest observed Milky Way satellites occupy halos with peak virial masses below $3.2\times 10^8 M_\odot$ at 95\% confidence. Furthermore, to reproduce the observed number of ultra-faint $M_V > -4$ Milky Way satellites, \citet{Jethwa.etal.2018}'s model favors a $>40\%$ occupation fraction for $M_{\rm vir} < 10^8 M_\odot$ halos. These findings appear to challenge our adopted 0\% occupation fraction cutoff at $3\times 10^8 M_\odot$, highlighting potential tensions between theoretical models and observational constraints.

Hydrodynamical simulations have reported a wide range of halo occupation fractions, reflecting the uncertainty in this relation. For instance, some studies have found that the peak halo mass at which 50\% of halos host galaxies of any mass, $M_{50}$, could reach values as high as $10^9 M_\odot$ \citep{Sawala.etal.2016, Fitts.etal.2017}, which aligns well with our adopted form (middle panel of Figure~\ref{fig:HMF_HOF_SMHM}). However, more recent high-resolution simulations of high-redshift galaxy formation, which incorporate the effects of spatially and temporally inhomogeneous reionization, suggest a lower threshold with $M_{50} \sim 10^8 M_\odot$ \citep{Katz.etal.2020}.

Moreover, recent studies have demonstrated that the halo occupation distribution is intricately linked to the halo's environment \citep{Bose.etal.2019, Hadzhiyska.etal.2020, Xu.etal.2021, Yuan.etal.2021, Hadzhiyska.etal.2023}. Specifically, these works reveal that, at a fixed halo mass, halos in denser environments tend to host a greater number of galaxies on average compared to those in underdense regions.

To ensure that our choice of halo occupation fraction does not significantly affect the computed volume filling factors, we also tested an alternative sigmoid function. This form assumes a 0\% occupation fraction for halos with peak virial masses below $10^7 M_\odot$ and a 100\% occupation fraction for halos above $5 \times 10^8 M_\odot$, consistent with the halo occupation fraction inferred from model fits to the DES satellite populations in \citet{Nadler.etal.2020}. The resulting volume filling factors differed by less than 5\%, indicating that our results are robust against shifts in the halo occupation fraction toward lower halo masses. As explained below, the specific form of the halo occupation fraction does not significantly impact our conclusions.

\subsection{Stellar mass -- halo mass relation (SMHM)}\label{sec:SMHM}

\begin{figure}
    \centering \includegraphics[width=\columnwidth]{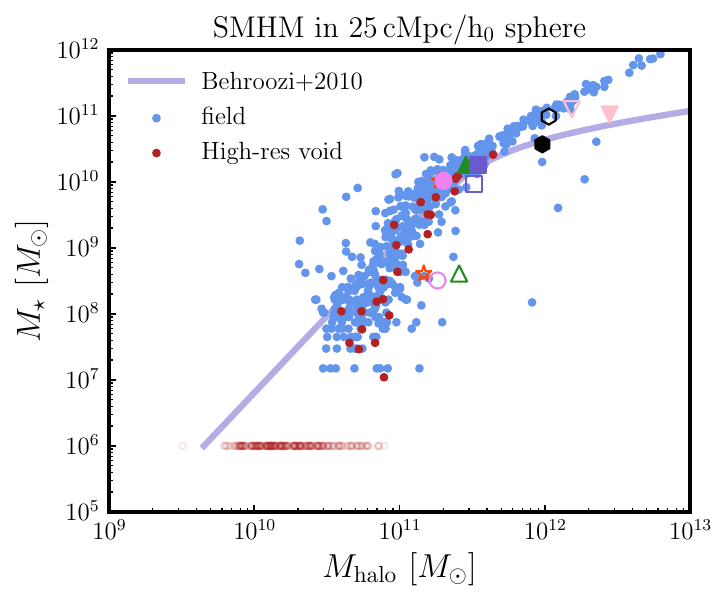}
    \caption{Stellar mass–halo mass relation (SMHM) for galaxies in the high-resolution void (red) and field (blue). We fit the SMHM data points using the model from \citet{Behroozi.etal.2010} and find excellent agreement for both field and void galaxies with $M_{\rm h} < 10^{11.5} \, M_\odot$. For more massive halos ($M_{\rm h} > 10^{11.5} \, M_\odot$), the void galaxies align well with theoretical predictions, while field galaxies deviate from this trend. Larger markers (pink circles, green triangles, purple squares, black hexagons, and light pink upside-down triangles) represent the six galaxies also present in the fiducial void simulation. Empty markers indicate galaxies in the fiducial void, while solid markers denote the same galaxies in the high-resolution run. Note that there are a total of 30 galaxies in the high-resolution void simulation, whereas only the six most massive galaxies are resolved in the fiducial run. Starless halos in both the field and high-resolution void are shown in the plot with empty red circles and are assigned a nominal $M_\star = 10^6 \, M_\odot$ to appear on the plot}. Due to the higher mass resolution of the high-resolution void compared to the field, we can resolve smaller halos in the high-resolution void run, although only those with $M_{\rm h} > 10^{10.3} \, M_\odot$ are capable of hosting galaxies.
    \label{fig:SMHM}
\end{figure}

The final step in obtaining a theoretical fit for the stellar mass function involves determining the stellar mass–halo mass relation. For this purpose, we adopt the widely utilized parameterization by \citet{Behroozi.etal.2010}, as implemented in \texttt{halotools} \citep{Hearin.etal.2017}. Various parameterizations of the stellar mass–halo mass relation exist in the literature, each derived from different modeling techniques. These include abundance matching approaches \citep[e.g.,][]{Moster.etal.2010, Guo.etal.2010, Wang.Jing.2010, Reddick.etal.2013, Moster.etal.2013}, halo occupation distribution and conditional luminosity function models \citep{Zheng.etal.2007, Yang.etal.2012}, as well as analyses based on cluster catalogs \citep{Yang.etal.2009, Hansen.etal.2009}. To assess the impact of variations in the stellar mass–halo mass relation on the final wind volume filling factor, we explore alternative forms of this relation in Sec.~\ref{sec:smhm_var}. This examination allows us to quantify the sensitivity of our results to the specific choice of the stellar mass–halo mass model.

Fig.~\ref{fig:SMHM} illustrates the stellar mass–halo mass relation for our simulated galaxies in the field, fiducial void, and high-resolution void simulations. Due to the low mass resolution of the fiducial void simulation, it contains only 6 galaxies with resolved stellar masses. In contrast, the high-resolution void simulation includes 30 galaxies, offering a more robust sampling. For halos with $M_{\rm h} < 10^{11.5} M_\odot$, both field and void galaxies closely follow the stellar mass–halo mass relation as parameterized by \citet{Behroozi.etal.2010}. Although we only implemented thermal supernova feedback and no momentum feedback, this agreement shows that at our simulation resolution, thermal feedback is sufficient. However, merely falling on this relation does not necessarily imply an accurate galaxy formation model, as the relation ignores the significant scatter, particularly at lower stellar masses. In the regime of more massive halos ($M_{\rm h} > 10^{11.5} M_\odot$), though, a notable divergence emerges: field galaxies exhibit higher stellar masses compared to the \citet{Behroozi.etal.2010} relation. This is probably due to AGN feedback at these massive scales being inaccurate in our simulation. However, we are only focusing on void galaxies in this project, and the most massive halo in a void is $\sim$ Milky Way-like. At our resolution, for the void galaxies, our recipe for thermal supernova feedback and AGN feedback is sufficient, and we postpone a more detailed study of how different feedback mechanisms could impact our results to future work.

The higher mass resolution of the high-resolution void simulation compared to the field simulation enables the resolution of smaller halos in the former. Specifically, halos with masses as low as $M_{\rm h} \approx 10^{9.5} M_\odot$ are resolved in the high-resolution void simulation, whereas the field simulation resolves halos only down to $M_{\rm h} \approx 10^{10} M_\odot$. However, despite this improved resolution, only the more massive halos with $M_{\rm h} > 10^{10.3} M_\odot$ are capable of hosting galaxies within the void environment.

To guide the eye, we use identical markers to represent the same six galaxies in both the fiducial and high-resolution void simulations, distinguishing them with empty and solid fills. Notably, for the two most massive halos (marked by a pink down triangle and a black hexagon) with $M_{\rm h} > 10^{12} M_\odot$, the differences in both halo mass and stellar mass between the two simulations are minimal. In contrast, among the four less massive halos with $M_{\rm h} < 10^{12} M_\odot$, three acquire significantly more stellar mass in the high-resolution simulation. This behavior suggests that the more massive halos are sufficiently resolved even in the low-resolution simulation, whereas the less massive halos are under-resolved. In the high-resolution simulation, the improved resolution allows for more star particles to be resolved, resulting in greater stellar mass for these smaller halos. This finding highlights the importance of running high-resolution simulations, and potentially even more resolved simulations in the future, to better capture the gas physics and star formation processes critical for accurately determining the stellar masses of low-mass halos. While infinite resolution would ideally provide the most accurate results, practical constraints such as computational resources and time make this infeasible. Consequently, a theoretical approach becomes essential. By combining the halo mass function, halo occupation fraction, and stellar mass–halo mass relation, we can derive a theoretical stellar mass function. Integrating this function within the simulation volume allows us to estimate the total number of galaxies under the assumption of infinite resolution. This, in turn, facilitates the calculation of the volume filling fraction by applying a wind model to the theoretical galaxy distribution.

By combining the halo mass function fitted for both the field and the high-resolution void simulation, the theoretical halo occupation fraction from \citet{Benitez-Llambay.Frenk.2020}, and the analytical stellar mass–halo mass relation from \citet{Behroozi.etal.2010}, we construct an analytical stellar mass function. This function is evaluated over 100 stellar mass bins to ensure sufficient resolution. To validate the robustness of our model, we compare the theoretical stellar mass function against the binned stellar mass functions derived from both simulations. The comparison, shown in the right panel of Figure~\ref{fig:HMF_HOF_SMHM}, provides a critical test of the agreement between the theoretical predictions and simulation results.

We begin by comparing the model stellar mass function to the simulation-derived stellar mass function for both the field and high-resolution void, accounting for the resolution limits of the simulations. The comparison reveals good agreement between the two, as shown by the blue solid line versus the purple dotted line for the field, and the red solid line versus the pink dotted line for the high-resolution void. Next, we extend the stellar mass function for both the field and high-resolution void simulations beyond the resolution limits. This extension spans from the filtering mass—the smallest halo capable of hosting a galaxy, assumed to be $3 \times 10^8 M_\odot$—to the most massive galaxy present in each simulation.

To calculate the total number of galaxies in both the resolution-limited and infinite-resolution cases, we integrate the theoretical stellar mass function. For the infinite-resolution case, the integration extends down to the filtering stellar mass limit, while for the resolution-limited case, it is restricted to the actual simulation resolution limit. In the resolution-limited case, we obtain a total of 303 galaxies in the field and 20 galaxies in the high-resolution void. These values are in reasonable agreement with the actual number of galaxies observed in the simulations, which are 542 for the field and 30 for the high-resolution void. For the infinite-resolution case, the integration yields significantly higher totals: 3981 galaxies in the field and 1099 galaxies in the high-resolution void. These results highlight the substantial number of dwarf galaxies lost due to the resolution limits of the simulations—thousands in the field and approximately 1000 in the void. This underscores the importance of accounting for resolution effects when interpreting simulation results and motivates the development of theoretical models to capture these unresolved populations.


\begin{figure*}
    \centering
    \includegraphics[width=\textwidth]{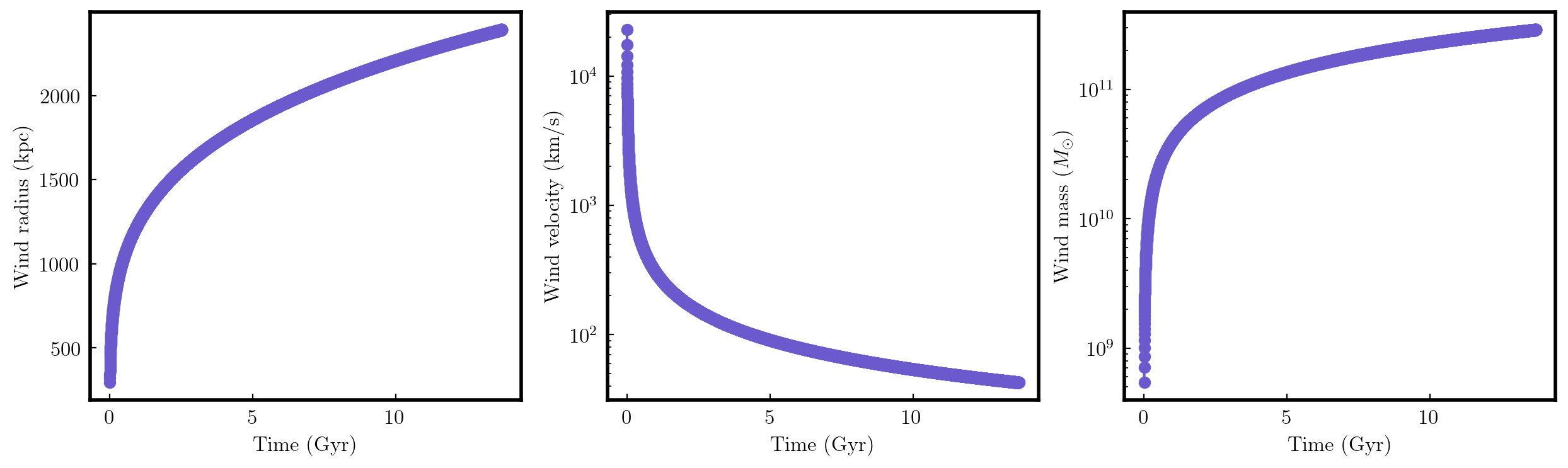}
    \caption{An example of the evolution of the wind radius (left), velocity (middle), and mass (right) for a sample galaxy with $M_\star = 3.33\times 10^{10} M_\odot$.} 
    \label{fig:wind_modeling}
\end{figure*}

\section{Galactic winds}\label{sec:wind}
Winds, or outflows, from galaxies churn out metals and pollute the inter-galactic medium (IGM) and circum-galactic medium (CGM) with metals. In the field (a normal environment with mean density equals $\Omega_{m,0} \rho_{c}$ and $\delta = 0$), galaxies pollute the IGM with metals, and we can calculate the fraction of volume these galaxies contaminate. For the void simulations, we can also calculate this volume filling fraction. This fraction should be much lower in void than the field, since we only have tiny dwarf galaxies and we have way less galaxies overall in the void. Although dwarf galaxies have much stronger outflows compared to more massive galaxies because of their shallower potential wells \citep{Dekel.Silk.1986, Efstathiou.1992, Mac-Low.etal.1999, Efstathiou.2000, Pontzen.etal.2012}, their small sizes and the low number of galaxies at all mass scales are the predominant factors so collectively the dwarf galaxies in the void can only contaminate a very small fraction of the entire volume. In this section, we use a simple supernova explosion model to explore the effect of outflows of dwarf galaxies below the simulation resolution limit in both environments on the volume filling fraction of galaxies.

\subsection{Wind modeling - a simple SNe explosion model}\label{sec:wind_modeling}

We adopt a simplified supernova explosion model for galactic outflows, following the approach of \citet{Efstathiou.2000} and \citet{Bertone.etal.2005} and originally proposed by \citet{Oort.1954}. In this model, supernovae deposit energy and impart momentum $P_{\rm SN}$ to drive winds. Assuming spherical symmetry and momentum conservation, the key governing equations are:
\begin{equation}
M = \frac{4}{3} \pi R^3 \rho_0, \quad Mv = P_{\rm SN}, \quad v = \dot{R},
\label{eqn:wind_eqn}
\end{equation}
where $\rho_0$ is the density of the intergalactic medium (IGM). The IGM density is linked to the critical density of the universe at $z=0$ as:
\begin{equation}
\rho_0 = \Omega_{\rm b} \rho_{\rm c},
\end{equation}
where we adopt $\Omega_{\rm b} = 0.04$ and $\rho_{\rm c} =3H_0^2/(8\pi G)$, with $H_0 = 67.81\ \rm km/s/Mpc$ \citep{Planck.2014}.

By solving Eq.~\ref{eqn:wind_eqn}, we find that the radius of the wind evolves as a function of time:
\begin{equation}
R(t) = \left( \frac{3P_{\rm SN}}{\pi\rho_0} t \right)^{1/4}.
\label{eqn:wind_rad}
\end{equation}
The wind expansion halts when its speed matches the IGM sound speed, $c_0$, which represents the level of turbulence in the IGM:
\begin{equation}
\dot{R} = c_0.
\label{eqn:wind_stop}
\end{equation}
Substituting Eq.~\ref{eqn:wind_rad} into Eq.~\ref{eqn:wind_stop}, we derive the wind stopping time:
\begin{equation}
t_{\rm stop} = (4c_0)^{-4/3}\left(\frac{3 P_{\rm SN}}{\pi \rho_0}\right)^{1/3}.
\end{equation}
The sound speed $c_0$ can be estimated using the formula for an ideal gas:
\begin{equation}
c_0 = \sqrt{\frac{\gamma k_{\rm B} T}{\mu m_{\rm p}}},
\end{equation}
where $\gamma = 5/3$ is the adiabatic index for a monatomic ideal gas, $T$ is the IGM temperature (assumed to be $\sim 10^4\ \rm K$), $\mu \sim 0.6$ is the mean molecular weight for ionized hydrogen, and $m_{\rm p}$ is the proton mass. Substituting these typical values, we obtain $c_0 \approx 2.63 \times 10^6\ \rm cm/s$.

For a typical supernova momentum injection of \(P_{\rm SN} \approx 10^{43}\ \rm kg \cdot m/s\) \citep{Kim.Ostriker.2015} and an intergalactic medium (IGM) density of \(\rho_0 \approx 1.67 \times 10^{-27}\ \rm kg/m^3\) \citep{Fukugita.etal.1998}, the stopping time for a single supernova explosion is estimated as:
\begin{equation}
t_{\rm stop} \approx 2.7\ \rm Gyr.
\end{equation}
The number of supernova explosions per galaxy, \(N_{\rm SN}\), can be approximated as \(N_{\rm SN} \sim M_\star / (100\ M_\odot)\). For the smallest galaxy in our void simulation (\(M_\star = 10^7 M_\odot\)), the shortest stopping time exceeds the age of the Universe. Consequently, we can safely integrate our model from the Big Bang to the present day. Using Eq.~\ref{eqn:wind_rad}, we calculate the final wind radius and the corresponding wind volume, representing the spatial extent of the wind at the present epoch.

The key parameter in this model is the total momentum deposited by supernovae into the wind, \(P_{\rm SN, tot}\), which is expressed as:
\begin{equation}
P_{\rm SN, tot} = N_{\rm SN} \cdot P_{\rm SN} \cdot K,
\end{equation}
where \(N_{\rm SN} \sim M_\star / (100\ M_\odot)\) represents the number of supernova explosions, \(P_{\rm SN}\) is the momentum injection per supernova (assumed to be \(\sim 10^5 M_\odot\ \rm km/s\); \citealt{Kim.Ostriker.2015}), and \(K\) characterizes the efficiency of momentum transfer into the wind.

In the following section, we apply the wind model to both field and high-resolution void galaxies. The parameter $K$ is determined by fitting the model to replicate the volume-filling fraction measured directly from gas cells in the simulations. Once calibrated, this $K$ is used to estimate the volume-filling fraction for the infinite-resolution case, providing a more accurate prediction.

Fig.~\ref{fig:wind_modeling} illustrates the evolution of wind properties for a galaxy with a stellar mass of $M_\star = 3.33 \times 10^{10} M_\odot$. The wind radius exhibits a fast increase up to $t_{\rm stop} \approx 5$ Gyr, after which its growth slows down, eventually reaching $R \sim 2500\ \rm kpc$ at the present day. Meanwhile, the wind velocity stabilizes at $\sim25$ km/s. This model underscores the dynamic interaction between supernova-driven momentum injection and IGM conditions in shaping galactic outflows.

\begin{figure}
    \centering
    \includegraphics[width=\columnwidth]{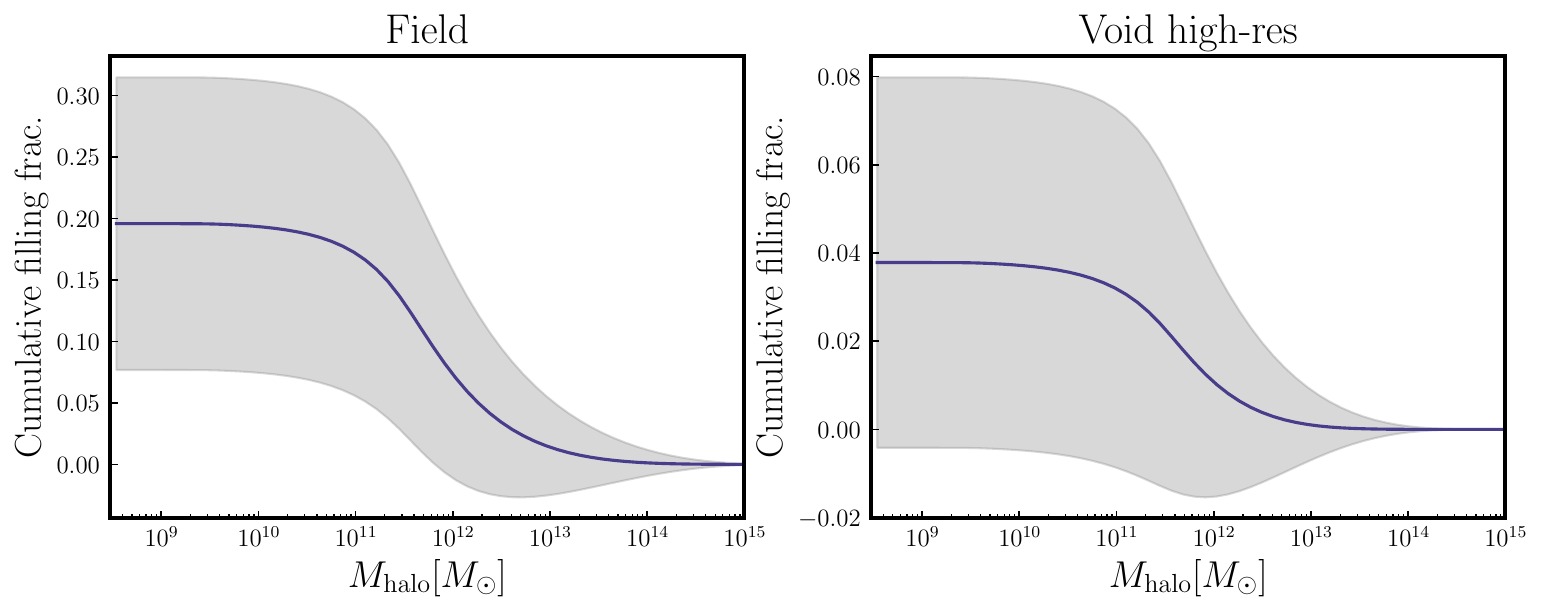}
    \caption{The cumulative volume contamination fraction as a function of halo mass, $M_{\rm h}$, is shown for both the field (left panel) and the void high-resolution simulation (right panel). Poisson error bars (shaded gray area) are propagated from the uncertainties of the halo mass function in the left panel of Figure~\ref{fig:HMF_HOF_SMHM}. Halos with $M_{\rm h} < 10^{10} \, M_\odot$ contribute negligibly to the contamination fraction. The majority of the contribution arises from halos in the range $10^{10} \, M_\odot < M_{\rm h} < 10^{12} \, M_\odot$.} 
    \label{fig:contam_frac_cum_Mh}
\end{figure}

\subsection{Volume filling fraction from wind modeling}\label{sec:Vf_model}
Galaxies drive outflows through supernova explosions, but the extent to which these outflows pollute the intergalactic medium (IGM) and the relative contributions of different types of galaxies remain open questions. Finite simulation resolution limits the capture of dwarf galaxies in halos below the resolution limit but above the "filtering mass"—the smallest halo mass capable of hosting a luminous galaxy \citep{Benitez-Llambay.Frenk.2020}. Using the wind modeling from Section~\ref{sec:wind_modeling}, combined with an analytical halo mass function, halo occupation fraction, and stellar mass--halo mass relation, we extend our analysis to include these lower-mass galaxies. This approach provides a more complete view of wind contamination in both void and field environments.

To compute the volume-filling fraction in resolution-limited simulations for both the field and high-resolution void, we first define a threshold for identifying "metal-rich" gas cells and calculate the volume-filling fraction based on their distribution. Further details on this procedure can be found in Section~\ref{sec:Vf_gas}. Using the wind model, which relies solely on galaxy stellar mass $M_\star$, we parameterize the results with the supernova efficiency $K$. By fitting $K$ to the observed volume-filling fractions from Section~\ref{sec:Vf_gas}, we derive the best-fit values for both field and void simulations, providing a consistent framework for comparing different environments.

Fig.~\ref{fig:contam_frac_K} shows the volume-filling fraction as a function of $K$ for both environments. As expected, higher supernova efficiency corresponds to larger volume-filling fractions due to increased energy driving galactic winds. Assuming spherical winds, the volume-filling fraction is computed by multiplying the wind volume of each galaxy by its abundance in each mass bin. Dividing this total volume by the simulation box volume (a spherical region of diameter $25\ \rm Mpc/h$) yields the wind-filling fraction. From Section~\ref{sec:Vf_gas}, the volume-filling fractions are $12.6\%$ for the field and $1.2\%$ for the void, computed from "metal-rich" gas cells in both simulations. 

We use our wind model to estimate the volume-filling fraction as a function of different $K$s. To incorporate uncertainties, the Poisson errors on the halo mass function are propagated to the wind fraction for different $K$ values, with the resulting errors shown as the grey shaded region in Figure~\ref{fig:contam_frac_K}. Horizontal dashed lines are drawn at the simulation volume-filling fractions of $12.6\%$ and $1.2\%$ to indicate the supernova efficiency values required to achieve these results. These efficiency values are estimated to be in the range of $0.1$--$0.3$.

It is important to note that supernova efficiency is primarily determined by the local properties of the galaxy's vicinity and the wind physics, making it independent of the large-scale environment. Consequently, the $K$ value is consistent across both void and field environments. In subsequent analyses, we adopt a representative supernova efficiency value of $0.2$, with an associated uncertainty of $0.1$.

With the supernova efficiency value $K=0.2$, the infinite-resolution wind-filling fraction is $18.6\%$, compared to $16.8\%$ for the resolution-limited case, indicating an additional $1.8\%$ contribution from dwarf galaxies in the halo mass range $10^6 M_\odot$ to $3\times10^8 M_\odot$. In void environments, the volume-filling fraction is $3.1\%$ for infinite simulation resolution and $2.9\%$ for limited resolution, highlighting a smaller contribution from dwarf galaxies in these regions.

As a consistency check, we estimate the volume filling fraction contributed by a single Milky Way-like galaxy, as illustrated in Figure~\ref{fig:wind_modeling}. According to the left panel of Fig.~\ref{fig:wind_modeling}, the final wind radius for such a galaxy is approximately $2$ Mpc, corresponding to a wind volume of $\sim 33\,\mathrm{Mpc}^3$. The number density of Milky Way-like galaxies is $\sim 10^{-2}/\mathrm{Mpc}^3$ in the field and $\sim 10^{-3}/\mathrm{Mpc}^3$ in void environments, as shown in the left panel of Figure~\ref{fig:HMF_HOF_SMHM}. Given these densities, the volume filling fraction contributed by a single Milky Way-like galaxy is $\sim 33\%$ in the field and $\sim 3.3\%$ in voids. These estimates are broadly consistent with the results obtained from our detailed numerical integration across all galaxy mass scales.

\section{Discussions}\label{sec:discussions}
In the previous sections, we presented our halo mass function + halo occupation fraction + stellar mass--halo mass relation model, along with the wind modeling, to fit the volume filling fraction derived from both the field and high-resolution void simulations. After determining the best-fit parameters for the model, we extended the halo mass range down to the smallest halo capable of hosting a galaxy (the "filtering mass") and up to the largest halo in either simulation. Using this extended range, we calculated the volume filling fraction for both simulations under the assumption of infinite resolution. Our results indicate that the contribution to the volume filling fraction from dwarf galaxies, which remain unresolved in the current simulations, is minimal.

\subsection{Volume filling fraction from dwarf galaxies }\label{sec:contribution_from_dwarfs}

In this section, we develop an analytical framework to quantify the contribution of dwarf galaxies to the volume contamination fraction. We begin by assuming a power-law form for the halo mass function of dwarf galaxies, as suggested by previous studies \citep{Jenkins.etal.2001, Reed.etal.2003, Vale.etal.2004, Santos-Santos.etal.2022}:
\begin{equation}
    \frac{dn}{dM_{\rm h}} \propto M_{\rm h}^a.
\end{equation}

The stellar mass--halo mass relation for dwarf galaxies can similarly be parameterized as a power law in the low-mass regime \citep{Brook.etal.2014, Read.etal.2017, Zaritsky.etal.2023, O'Leary.etal.2023}:
\begin{equation}
    M_\star \propto M_{\rm h}^b.
\end{equation}

From our wind modeling, the volume of each wind for a galaxy can be expressed as:
\begin{equation}
    V \propto M_\star^{3/4} = M_{\rm h}^{3b/4}.
\end{equation}

To compute the volume fraction per mass bin, we multiply the halo mass function, $\frac{dn}{dM_{\rm h}}$, by the volume, $V$, associated with each $M_{\rm h}$:
\begin{equation}
    \frac{df}{dM_{\rm h}} \propto M_{\rm h}^{a + \frac{3b}{4}}.
\end{equation}

The values of $a$ and $b$ are critical in determining the behavior of $\frac{df}{dM_{\rm h}}$. Using the \citet{Tinker.etal.2008} relation, we find $\frac{dn}{dM_{\rm h}} \propto M_{\rm h}^{-1.9}$. Alternatively, the Press-Schechter halo mass function \citep{Press.Schechter.1974} provides:
\begin{align}
    \frac{dn}{dM_{\rm h}} &\propto \frac{\bar{\rho}}{M_{\rm h}^2} \left( \frac{M_{\rm h}}{M^*} \right)^{(3+n)/6} \exp \left( - \left( \frac{M_{\rm h}}{M^*} \right)^{(3+n)/3} \right),
\end{align}
where $n$ is the index of the power spectrum of primordial fluctuations, $P(k) \propto k^n$, $\bar{\rho}$ is the mean matter density, and $M^*$ is the characteristic cutoff mass. In the dwarf galaxy regime, $n \approx -2$, as suggested by studies analyzing HI intensity fluctuations. For example, \citet{Dutta.etal.2009} estimated the power spectrum of HI intensity fluctuations in dwarf galaxies, finding slopes in the range of approximately $-1.5$ to $-2.6$. This and similar results support the approximation of $n \approx -2$ and justify neglecting the exponential cutoff in the Press-Schechter formalism for this regime.

However, it is important to note that the exact value of $n$ on these scales remains uncertain. For instance, \citet{Dekker.Kravtsov.2024} explored the effects of a tilted primordial power spectrum on the properties of dwarf galaxies, finding that variations in $n$ on small scales can significantly impact galaxy formation. Similarly, \citet{Yoshiura.etal.2020} constrained the small-scale primordial power spectrum using galaxy UV luminosity functions, probing wavenumbers as high as $k \sim 10^3 \, \mathrm{Mpc}^{-1}$. These studies suggest that while $n \approx -2$ is a reasonable approximation for the dwarf galaxy regime, further investigations into small-scale deviations from scale invariance are necessary to fully understand the role of primordial fluctuations in shaping dwarf galaxies.

Finally, for Press-Schechter with the above power-law index in the dwarf galaxies regime, the halo mass function becomes:
\begin{equation}
    \frac{dn}{dM_{\rm h}} \propto M_{\rm h}^{-11/6}.
\end{equation}

This yields $a = -11/6$, which is close to the \citet{Tinker.etal.2008} value of $-1.9$. For the stellar mass--halo mass relation, using the \citet{Behroozi.etal.2010} model, we find $b \approx 2.32$. Combining these results, the volume contamination fraction per mass bin is given by:
\begin{equation}
    \frac{df}{dM_{\rm h}} \propto M_{\rm h}^{a + \frac{3b}{4}} \propto M_{\rm h}^{-0.16} \quad \text{or} \quad M_{\rm h}^{-0.09}.
\end{equation}

Upon integration, the cumulative contamination fraction is:
\begin{equation}
    f \propto \int \frac{df}{dM_{\rm h}} \, dM_{\rm h} \propto M_{\rm h}^{0.84} \quad \text{or} \quad M_{\rm h}^{0.91}.
\end{equation}

This result converges for decreasing $M_{\rm h}$, consistent with the plateau behavior observed in Figure~\ref{fig:contam_frac_cum_Mh}. It confirms that the contribution of dwarf galaxies to the volume contamination fraction is minimal and decreases as $M_{\rm h}^{0.8 \sim 0.9}$.

\subsection{Cosmic variance of halo mass function}\label{sec:cosmic_variance}

\begin{figure*}
    \centering
    \includegraphics[width=\textwidth]{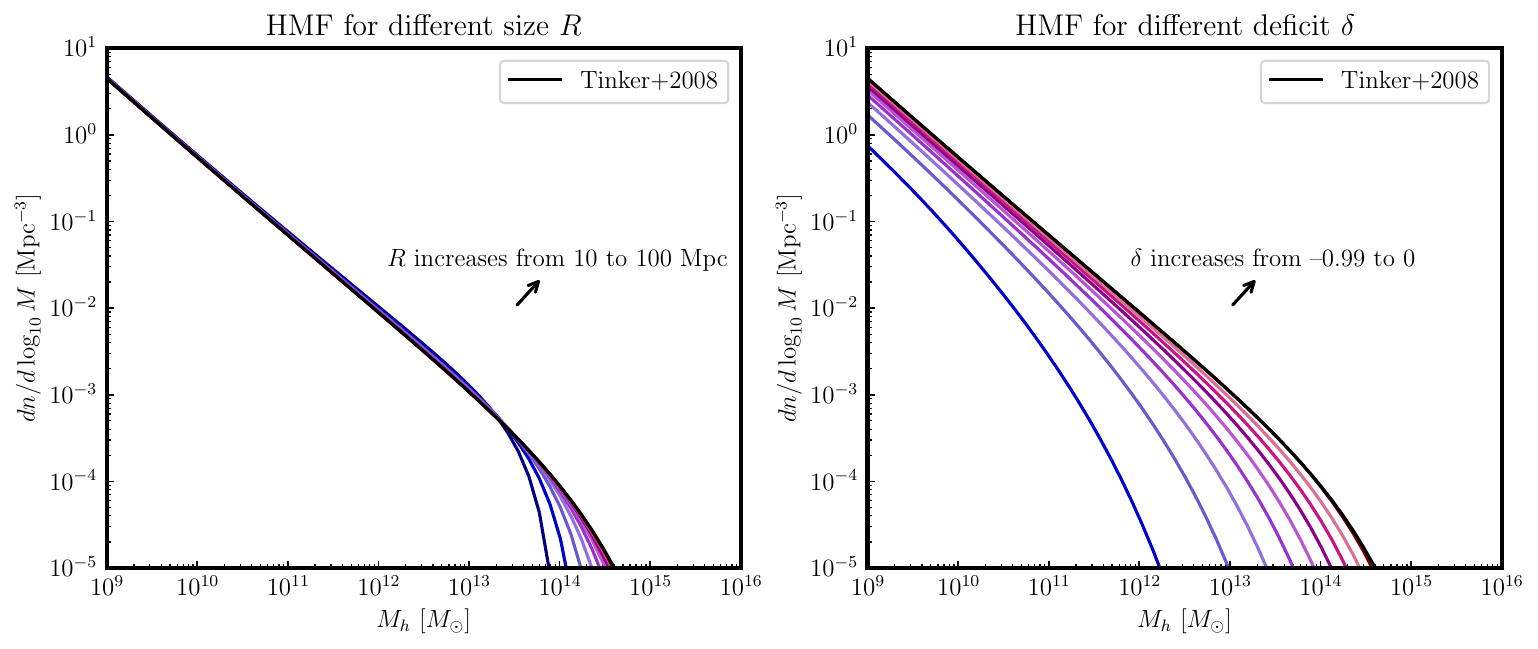}
    \caption{\emph{Left:} Halo mass function for various sizes $R$ ranging from 10 Mpc to 100 Mpc, at fixed $\delta = 0$. The curves correspond to increasing $R$, progressing from the bottom to the top. For comparison, the results from \citet{Tinker.etal.2008} are overplotted. As $R$ increases, the Halo mass function  gradually converges toward the infinite $R$ case represented by \citet{Tinker.etal.2008}. At small $R$, the abundance of large halos with $M_{\rm h} > 10^{13.5} M_\odot$ is truncated due to the inability of massive halos to form within finite simulation boxes. \emph{Right:} Halo mass function for varying environmental deficits $\delta$, ranging from $-0.99$ to $0$, at fixed $R = 60$ Mpc. The curves correspond to increasing $\delta$, progressing from the bottom to the top. In increasingly underdense environments, the number of halos across all mass scales is suppressed, with the suppression being more pronounced for larger halos.
} 
    \label{fig:HMF_size_delta}
\end{figure*}

In this section, we explore how our results change for different halo mass functions, taking into account variations in void properties such as density deficit ($\delta$) and size ($R$). This analysis is motivated by the fact that we investigate only a single void in this paper. In contrast, simulations like the CAMELS project\footnote{\href{}{https://www.camel-simulations.org/}} \citep{CAMELS_presentation, CAMELS_DR1, CAMELS_DR2} and observational datasets such as the Sloan Digital Sky Survey \citep[SDSS,][]{Mao.etal.2017.void, Douglass.etal.2023} contain thousands of voids with diverse density deficits and sizes. In each void, the halo mass function can vary significantly, influencing the theoretical void volume filling fraction. To address this, we examine how changes in $\delta$ and $R$ impact the halo mass function and, subsequently, the volume filling fraction.

Our void, characterized by a density deficit of $\delta = -0.8$, results in a volume filling fraction of $3.1\%$ at infinite resolution for $K=0.2$. For $\delta = -0.6$ and $\delta = -0.9$, the corresponding volume filling fractions are $6.4\%$ and $1.5\%$, respectively. These trends are consistent with the expectation that a higher $\delta$ (indicating less underdensity) implies more matter within the void and, consequently, greater filling from galaxies.

Figure~\ref{fig:HMF_size_delta} illustrates the impact of void properties on the halo mass function by showing the dependence of the halo mass function on the size $R$ (left panel) and density deficit $\delta$ (right panel). In the left panel, the halo mass function is plotted for increasing size, $R$, ranging from $10$ to $100$ Mpc at a fixed deficit value of $\delta = 0$. As $R$ increases, the halo mass function gradually converges to the present-day \citet{Tinker.etal.2008} halo mass function for infinite $R$ and $\delta = 0$. For smaller sizes, or equivalently smaller box sizes, the absence of more massive halos is evident because smaller boxes cannot accommodate large halos.

The right panel of Figure~\ref{fig:HMF_size_delta} shows the halo mass function for increasing $\delta$, ranging from $\delta = -0.99$ to $\delta = 0$. The halo mass function converges to the Tinker halo mass function as $\delta$ approaches the field value ($\delta = 0$). At lower $\delta$ values, the number density of halos across all mass scales is suppressed, with the suppression being more pronounced for massive halos compared to smaller ones. This behavior highlights the significant impact of $\delta$ on the halo mass function.

In Section~\ref{sec:vff_delta_size}, we further explore how these variations in the halo mass function, driven by changes in $\delta$ and $R$, affect the theoretical void volume filling fraction. These results underline the importance of void properties in shaping the halo mass function and the consequent impact on galaxy distribution within voids.


\subsection{Volume filling fraction as a function of deficit and size}\label{sec:vff_delta_size}

\begin{figure}
    \centering
    \includegraphics[width=\columnwidth]{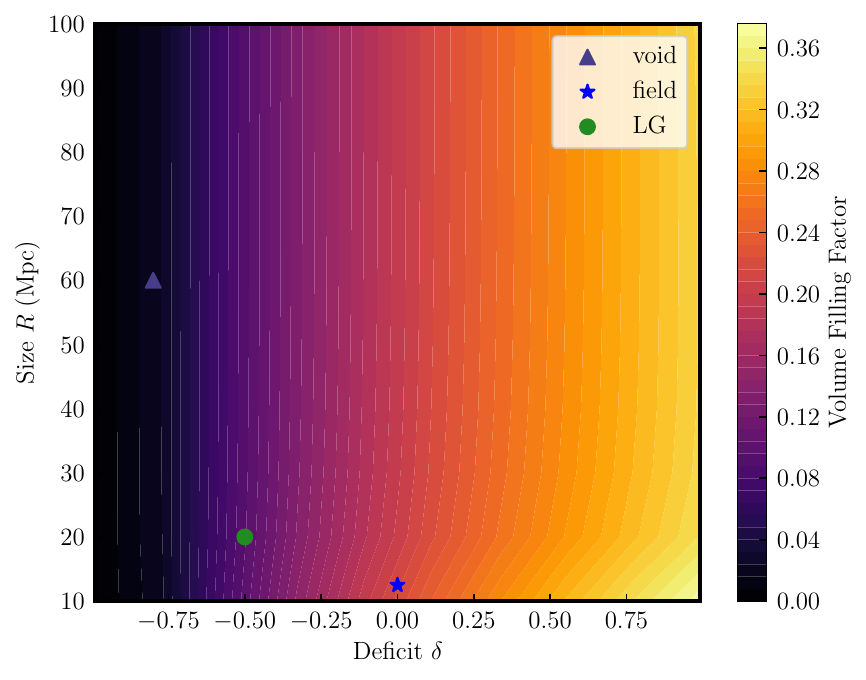}
    \caption{Volume filling fraction as a function of size $R$ and environmental deficit $\delta$ for supernova momentum injection efficiency of $K=0.2$. We show the volume filling fraction derived using our void and field simulations as purple triangle and blue star, respectively, and for the LG condition (Section~\ref{sec:MW_void}) as green circle. The results show that the volume filling fraction is nearly independent of $R$, except 
    for $R < 20$ Mpc.} 
    \label{fig:vff_delta_size}
\end{figure}

Figure~\ref{fig:vff_delta_size} illustrates the volume filling factor as a function of both size, $R$, and density deficit, $\delta$, resulting from variations in the halo mass function (discussed in Section~\ref{sec:cosmic_variance}). In this analysis, we take the supernova momentum injection efficiency to be $K = 0.2$.

An intriguing result emerges: except in environments that are nearly field-like ($\delta \approx 0$), the volume filling factor is largely independent of the size $R$. However, it exhibits a strong dependence on $\delta$. For the smallest sizes ($R < 20 \, \mathrm{Mpc}$), the volume filling factor increases with both $R$ and $\delta$. This behavior highlights the dominant influence of the density deficit on the filling factor in void-like environments, while the size primarily affects the smallest-scale structures.


\subsection{HMF and SMHM variations}\label{sec:smhm_var}
Except for the halo mass function and wind modeling, other aspects of our model that could influence the final wind contamination fraction include variations in the halo occupation fraction and the stellar mass–halo mass relation. In this section, we analyze how differences in these factors affect the wind contamination fraction across diverse environments, highlighting the robustness of our results to variations in halo occupation fraction.

Recent studies, such as \citet{Nadler.etal.2020}, offer alternative halo occupation fraction models derived from statistical inference for satellite galaxies, which differ from the field halo occupation fraction used in our model. Their model shows a lower cutoff mass for satellite galaxies, reflecting physical processes such as host halo retention of gas reservoirs against supernova feedback and tidal stripping that reduces satellite halo mass while preserving stellar mass. Additionally, satellite galaxies forming stars pre-reionization may have a lower cutoff mass compared to field galaxies, which predominantly form post-reionization. Despite these differences, overlaying the halo occupation fraction from \citet{Nadler.etal.2020} with our model reveals minimal variation in the predicted wind contamination fraction. This indicates that the wind contamination fraction is relatively insensitive to the specific choice of halo occupation fraction model, even when accounting for significant differences in galaxy formation processes between field and satellite environments.

This robustness underscores the broader conclusion that the final wind contamination fraction is primarily driven by the halo mass function and wind modeling, with variations in halo occupation fraction contributing only minor deviations. The small sensitivity to halo occupation fraction differences suggests that our model captures the dominant physical mechanisms regulating wind contamination across environments. Incorporating error bars derived from Poisson statistics further supports this conclusion, as the uncertainties remain well within the range of variation observed between different halo occupation fraction models. This finding emphasizes the utility of our approach for studying galactic winds and their environmental dependence, despite ongoing debates about the exact nature of halo occupation fraction in different settings.

The stellar mass–halo mass relation represents another critical component of our model, and uncertainties at the low-mass end introduce additional complexities. Studies such as \citet{Feldmann.etal.2019} and \citet{Monzon.etal.2024} highlight significant scatter in the stellar mass–halo mass relation for low-mass halos, driven by factors such as feedback from supernovae and the impact of reionization. For instance, \citet{Kim.etal.2024} emphasize that the stellar mass in these halos is highly sensitive to the efficiency of feedback processes, leading to a broad range of predicted stellar-to-halo mass ratios. Similarly, \citet{Sawala.etal.2015} find that halos with masses below $10^{10} M_\odot$ are particularly affected by reionization, which suppresses gas accretion and star formation, resulting in a steep decline in the stellar mass–halo mass relation.

Additional complexities arise from observational constraints, such as those discussed in \citet{Danieli.etal.2023} and \citet{Rohr.etal.2022}, which suggest that low-mass galaxies exhibit greater diversity in their star formation histories and stellar masses compared to their high-mass counterparts. This diversity complicates efforts to define a universal stellar mass–halo mass relation at the low-mass end. Furthermore, \citet{Girelli.etal.2020} and \citet{Hadzhiyska.etal.2023} note that the scatter in the stellar mass–halo mass relation becomes increasingly pronounced in underdense environments, such as cosmic voids, where the formation of low-mass galaxies is further influenced by their isolation and unique feedback mechanisms.

To refine our analysis, we introduced a scaling factor, $(M_\star/M_{\rm h})_0$, to the stellar mass–halo mass relation from \citet{Behroozi.etal.2010} and used \texttt{scipy.minimize} to optimize $(M_\star/M_{\rm h})_0$ for the best fit to the stellar mass–halo mass relations derived from both field and high-resolution void simulations. Our optimization yielded $(M_\star/M_{\rm h})_0 = 0.97$ for void environments and $(M_\star/M_{\rm h})_0 = 1.01$ for the field, indicating that, on average, field galaxies have slightly higher stellar masses than their void counterparts at fixed halo mass. This result is consistent with the expectation that galaxies in underdense environments experience weaker gas accretion and feedback effects, leading to reduced star formation compared to galaxies in denser regions. The relatively small variation in $(M_\star/M_{\rm h})_0$ suggests that environmental effects on the stellar mass–halo mass relation are modest.

To assess the impact of different stellar mass–halo mass relations on our results, we explored two extreme cases for $(M_\star/M_{\rm h})_0$. First, we set $(M_\star/M_{\rm h})_0 = 2$, representing an enhanced stellar mass–halo mass relation, as might be expected in high-gas environments with increased star formation efficiency. In this scenario, the final wind volume filling fraction at infinite resolution increases to 34.2\% in the field and 5.2\% in voids, compared to 18.6\% and 3.1\%, respectively, for the original Behroozi stellar mass–halo mass relation. Conversely, for a low-star formation case with $(M_\star/M_{\rm h})_0 = 0.2$, the volume filling fraction drops to 6.1\% in the field and 0.9\% in voids. These results demonstrate that while the stellar mass–halo mass relation remains approximately consistent between voids and the field in our simulations, extreme variations in this relation can significantly impact the predicted volume filling fractions.

\subsection{Milky Way in a void}\label{sec:MW_void}
Our model provides a versatile framework for predicting the wind contamination fraction across a wide range of environments, including the Local Group and Milky Way. The Simulating the LOcal Web (SLOW) simulation presented in \citet{Dolag.etal.2023} offers critical insights into the local density field, particularly through its cumulative relative density profiles for dark matter and galaxies. The void density profile shown in Figure 5 of that work enables us to identify key parameters, such as the void size $R$ and the density deficit $\delta$, that are essential for applying our model to the Milky Way environment.

From the SLOW simulation, the local void surrounding the Milky Way exhibits a density deficit of $\delta \approx -0.5$ and a size extending to approximately $R\approx$ 20 Mpc, characterized by a steep rise in density at its edge. Incorporating these values into our model allows us to account for the Milky Way's underdense environment when calculating the wind contamination fraction. By integrating the void's density profile with our halo occupation fraction and stellar mass--halo mass relation, our framework predicts a volume filling fraction of $9.6\% \pm 3.3\%$ for the Milky Way environment.

The ability to use void parameters directly from simulations such as SLOW underscores the flexibility of our approach. Whether in cosmic voids with extreme underdensities or the more complex environment of the Local Group, the core components of our model – the halo mass function, halo occupation fraction, stellar mass--halo mass relation, and wind modeling – remain applicable. For the Milky Way, this means we can predict how local variations in galaxy density and stellar feedback impact the wind contamination fraction, while leveraging observational constraints and simulation results to refine our predictions. This adaptability highlights the power of combining theoretical models with simulation-based density profiles to study galactic winds across diverse cosmic environments.

\section{Summary}\label{sec:summary}

We present a theoretical framework to calculate the volume filling fraction of galaxies, combining analytical models such as the halo mass function, halo occupation fraction, stellar mass--halo mass relation, and galactic wind modeling. Using \texttt{RAMSES}, we perform a zoom-in hydrodynamical simulation on a $25\, \mathrm{cMpc}/\rm{h_0}$ spherical void, identified as the most underdense region from 1,000 randomly selected spheres, and compare its properties with a non-zoom-in field simulation of the same size and initial conditions. The void, with a density deficit of $\delta \approx -0.8$ and an actual size of $\sim$120$\, \mathrm{cMpc}/\rm{h_0}$, demonstrates that galaxy properties within the central $25\, \mathrm{cMpc}/\rm{h_0}$ region remain unchanged when the zoom volume is expanded, though higher resolution improves results.

However, accurately capturing the physics of galaxy formation within voids requires sufficient mass and spatial resolution. The impact of simulation resolution effects is significant: low-resolution simulations under-resolve low-mass halos, leading to underestimated stellar masses. In contrast, high-resolution simulations allow more star particles to be resolved, improving the accuracy of stellar mass predictions and better capturing the physics of gas dynamics and star formation. Despite these advantages, practical computational constraints prevent arbitrarily high resolution. To address this, we develop a theoretical framework that extrapolates simulation results to infinite resolution, ensuring robust modeling of star formation and galactic outflows.

Our high-resolution simulation further validates the analytical void halo mass function, demonstrating strong agreement between the simulation-derived mass function and theoretical models that account for void underdensity and limited box size. By integrating the halo mass function, a sigmoid-model halo occupation fraction, and the stellar mass--halo mass relation with supernova-driven wind modeling, we predict the volume filling fraction for both field and void environments under the assumption of infinite resolution. We estimate filling fractions of 18.6\% in the field and 3.1\% in the void, with variations depending on void size and $\delta$. Our framework highlights that resolving halos down to $M_{\rm h} \sim 10^{10} M_\odot$ is sufficient for reliable filling fraction predictions, reducing computational requirements while remaining robust across diverse environments.

\begin{itemize}

    \item Galaxy properties in the central $25\, \mathrm{cMpc}/\rm{h_0}$ region of the void remain unchanged when the zoom-in volume is increased, but higher resolution significantly improves the accuracy of the physics of galaxies.
    
    \item The high-resolution void simulation shows strong agreement between the halo mass function derived from the simulation and analytical predictions from extended Press-Schechter that account for void underdensity and box size limitations.
    
    \item The volume filling fraction is predicted to be 18.6\% in the field and 3.1\% in the void with a density deficit of $\delta = -0.8$ assuming infinite resolution in the simulation, and it varies depending on void properties such as size and underdensity, although the dependence on size is much weaker than density deficit.
    
    \item Cosmic variance studies reveal substantial scatter in volume filling fraction predictions, driven by variations in void size and density contrast, emphasizing the need for a statistical approach when interpreting results across void samples.

    \item Dwarf galaxies are found to contribute to the volume contamination fraction following a scaling relation of $M_{\rm h}^{0.8-0.9}$.
    
    \item Resolving halos down to a mass scale of $M_{\rm h} \sim 10^{10} M_\odot$ is sufficient for accurately calculating theoretical volume filling fractions, ensuring robust predictions without requiring simulations to resolve down to the filtering mass.

    \item The applicability of our framework to real-world systems, such as the Local Group, demonstrates its versatility. For example, our predictions align well with observations for the Milky Way's void, characterized by $\delta \approx -0.5$ and a radius of $\sim20\, \mathrm{Mpc}$.

\end{itemize}

\section*{Acknowledgements}
The simulations in this study were conducted using computational resources provided by Princeton Research Computing, which is supported by the Princeton Institute for Computational Science and Engineering (PICSciE) and the Office of Information Technology’s High Performance Computing Center and Visualization Laboratory at Princeton University.

AP acknowledges support from the European Research Council (ERC) under the European Union’s Horizon programme (COSMOBEST ERC funded project, grant agreement 101078174), as well as support from the french government under the France 2030 investment plan, as part of the Initiative d’Excellence d’Aix-Marseille Université - A*MIDEX AMX-22-CEI-03.

\section*{Data Availability}
The simulations were run on the Stellar cluster at Princeton University. Access to this data is available upon request to the corresponding author.


\bibliographystyle{mnras}
\bibliography{main} 

\appendix

\section{Volume filling fraction from simulation}\label{sec:Vf_gas}
\begin{figure}
    \centering
    \includegraphics[width=\columnwidth]{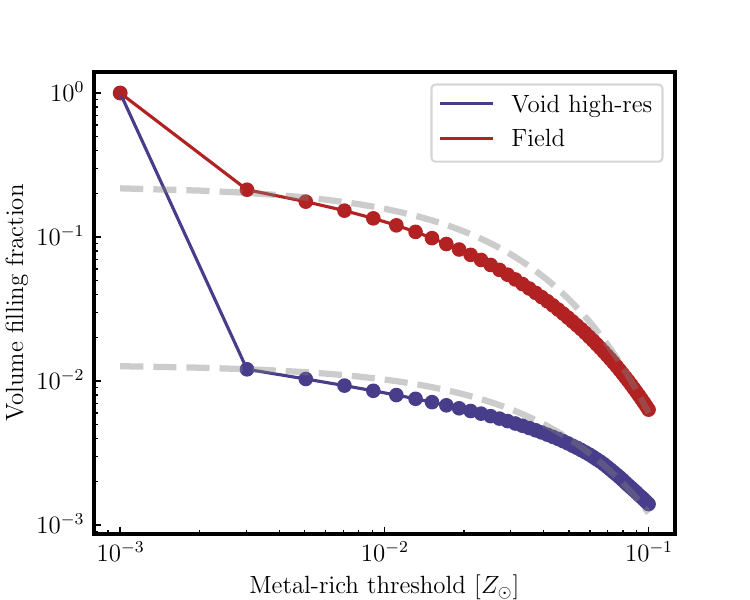}
    \caption{Volume filling fraction as a function of the metallicity threshold (in units of solar, $Z_\odot = 0.02$) we use to determine if a gas cell is metal-rich or not. As we decrease the threshold value, more gas cells are considered as "metal-rich". When we decrease to the minimum metallicity value we start the simulation with ($10^{-3}Z_\odot$), since every cell has a metallicity higher than this value we have a 100\% volume filling fraction. We fit a flat line for the low metal-rich threshold regime and a power law for the high metal-rich threshold regime to both the void high-resolution and field simulations (grey dashed curves). The trend plateaus to $\sim 12.6\%, 1.2\%$ for smaller metal-rich threshold, and we take these two values to be the converged, resolution-limited volume filling fraction for the field and void high-res, respectively.  
    } 
    \label{fig:contam_frac_gas_cells}
\end{figure}

To compute the volume-filling fraction from gas cells in the simulation, a metallicity threshold is required to identify "metal-rich" gas cells – those that are considered contaminated by galactic winds. Figure~\ref{fig:contam_frac_gas_cells} illustrates the volume-filling fraction as a function of the metallicity threshold, expressed in units of solar metallicity ($Z_\odot = 0.02$). For this calculation, a range of metallicity thresholds is applied, and any gas cell with a metallicity exceeding the given threshold is flagged as contaminated. The total volume of all flagged metal-rich gas cells is then summed and normalized by the box volume to determine the volume-filling fraction.

As the metallicity threshold decreases, an increasing number of gas cells are classified as "metal-rich." At the lowest metallicity value used to initialize the simulation ($10^{-5}$, or $0.1\%\ Z_\odot$), all gas cells exceed this threshold, resulting in a volume-filling fraction of 100\%. In the low-threshold regime, the volume filling factor should be flat, as it reaches a maximum value once all enriched bubbles are included. We therefore fit a two-part function—a flat line at low thresholds and a power-law decline at high thresholds—to the high-resolution void and field simulations (purple and red curves in Figure~\ref{fig:contam_frac_gas_cells}). The absence of a sharp transition between the flat and power-law regions in Figure~\ref{fig:contam_frac_gas_cells} arises from numerical diffusion at the edges of the bubbles. This diffusion smooths the transition between enriched and pristine gas, making it difficult to model the effect precisely. As such, we adopt a visual fit to approximate the two regimes. Although the exact convergence point is difficult to quantify, this uncertainty does not affect the scientific conclusions of our model. It merely alters the resolution-limited value of the simulated volume filling factor, which functions as a tunable parameter in practical applications of our framework.

For lower threshold values, the volume-filling fraction stabilizes, plateauing at approximately $12.6\%$ for the field and $1.2\%$ for void high-resolution simulations. These plateau values represent the resolution-limited, converged volume-filling fractions for each simulation.

The comparison reveals that the volume-filling fraction is 10\% lower in void environments than in the field. This significant difference reflects the paucity of galaxies in voids, underscoring the impact of environmental factors on the distribution of metal-enriched gas.

\begin{figure}
    \centering
    \includegraphics[width=\columnwidth]{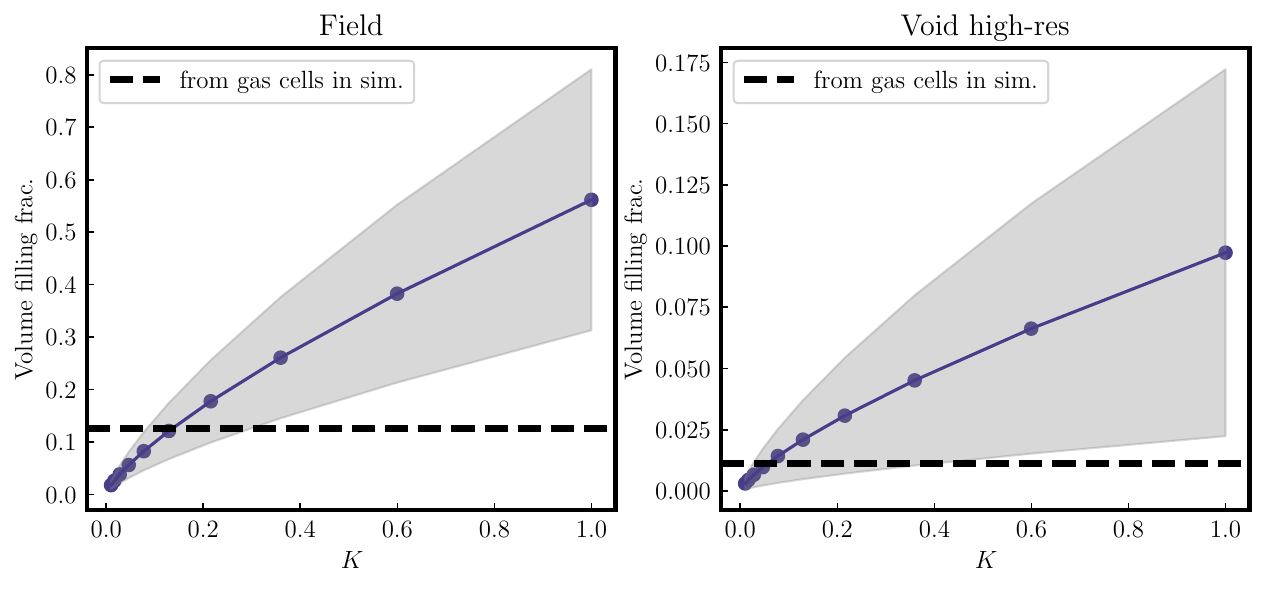}
    \caption{Volume-filling fraction as a function of supernova efficiency $K$ for both the field (left) and the high-resolution void (right), computed from our analytical model. The black horizontal dashed line represents the volume-filling fraction derived in Section~\ref{sec:wind_modeling} using gas cells. As anticipated, increasing the supernova efficiency results in greater energy input into galactic winds, thereby producing a higher volume-filling fraction. The supernova efficiency required to reproduce the observed volume-filling fractions for the field and the high-resolution void simulations is $0.1-0.3$ and $0.05-0.4$, respectively. Since $K$ is a local quantity related to wind physics and independent of large-scale environment, it should be the same in void and field. Thus, we have $K=0.2$ with an associated uncertainty of 0.1.} 
    \label{fig:contam_frac_K}
\end{figure}

\bsp
\label{lastpage}
\end{document}